\documentclass[conference]{IEEEtran}
\usepackage[utf8]{inputenc}
\usepackage[T1]{fontenc}
\usepackage{amsmath}
\usepackage{amssymb}
\usepackage{graphicx}
\usepackage{hyperref}
\usepackage{float}
\usepackage{booktabs}
\usepackage{cuted}

\usepackage[table]{xcolor}
\renewcommand{\textcolor}[2]{#2}

\begin{document}

\title{A Survey of Heterogeneous Graph Neural Networks for Cybersecurity Anomaly Detection}

\author{\IEEEauthorblockN{Laura Jiang, Reza Ryan, Qian Li, Nasim Ferdosian}
\IEEEauthorblockA{School of Electrical Engineering, Computing and Mathematical Sciences\\
Curtin University, Perth, Australia}}

\maketitle

\begin{abstract}
Anomaly detection is a critical task in cybersecurity, where identifying insider threats, access violations, and coordinated attacks is essential for ensuring system resilience. Graph-based approaches have become increasingly important for modeling entity interactions, yet most rely on homogeneous and static structures, which limits their ability to capture the heterogeneity and temporal evolution of real-world environments. Heterogeneous Graph Neural Networks (HGNNs) have emerged as a promising paradigm for anomaly detection by incorporating type-aware transformations and relation-sensitive aggregation, enabling more expressive modeling of complex cyber data. However, current research on HGNN-based anomaly detection remains fragmented, with diverse modeling strategies, limited comparative evaluation, and an absence of standardized benchmarks. To address this gap, we provide a comprehensive survey of HGNN-based anomaly detection methods in cybersecurity. We introduce a taxonomy that classifies approaches by anomaly type and graph dynamics, analyze representative models, and map them to key cybersecurity applications. We also review commonly used benchmark datasets and evaluation metrics, highlighting their strengths and limitations. Finally, we identify key open challenges related to modeling, data, and deployment, and outline promising directions for future research. This survey aims to establish a structured foundation for advancing HGNN-based anomaly detection toward scalable, interpretable, and practically deployable solutions.
\end{abstract}


\paragraph{Abbreviations:}
\begin{itemize}
    \item APT: Advanced Persistent Threat
    \item AUPRC: Area Under the Precision--Recall Curve
    \item AUROC: Area Under the Receiver Operating Characteristic Curve
    \item GAT: Graph Attention Network
    \item GCN: Graph Convolutional Network
    \item GNN: Graph Neural Network
    \item GRU: Gated Recurrent Unit
    \item HGNN: Heterogeneous Graph Neural Network
    \item LSTM: Long Short-Term Memory
    \item SOC: Security Operations Center
\end{itemize}

\section{Introduction}
Anomaly detection is fundamental to maintaining the security and resilience of cyber systems. In cybersecurity, anomalies often correspond to access violations, insider threats, or coordinated attacks that deviate from expected system behavior. As digital environments grow in complexity, traditional flat-feature modeling becomes insufficient. Instead, graph-structured data has emerged as a natural and powerful way to represent relational interactions among entities such as users, hosts, files, and processes. \cite{1}

Cybersecurity graphs are typically heterogeneous and dynamic, consisting of multiple entity types connected via diverse relations that evolve over time. \cite{2} These structural and temporal complexities present significant challenges for anomaly detection models. While Graph Neural Networks (GNNs) have become a standard approach for learning from graph data, most existing methods assume homogeneous and static structures. \cite{3} This limits their effectiveness in real-world cyber scenarios, where capturing semantic heterogeneity and behavioral evolution is essential. To address this gap, Heterogeneous Graph Neural Networks (HGNNs) introduce type-aware transformations and relation-sensitive aggregation mechanisms \cite{4}, enabling more accurate modeling of the underlying semantics in such multi-typed environments. Although HGNNs have been increasingly applied to anomaly detection in cybersecurity, the field remains fragmented, with diverse modeling strategies and limited comparative analysis, highlighting the need for a structured survey.

Moreover, most existing surveys \cite{5,6,7,8} focus on homogeneous or static graphs, offering limited insight into models tailored for heterogeneous and dynamic graphs. In this survey, we bridge this gap by reviewing recent advances in HGNN-based anomaly detection methods, with a particular emphasis on their applicability to cybersecurity. Recent cybersecurity-specific graph anomaly detection studies further underscore the pace of change in this area, including self-supervised intrusion detection systems such as Anomal-E \cite{9} and TS-IDS \cite{10}, graph-based log anomaly detection frameworks with explicit explanation components \cite{11}, interpretable spatio-temporal log models such as IST-GCN \cite{12}, and 2025 intrusion-detection extensions including BS-GAT \cite{13}, TE-G-SAGE \cite{14}, and real-world IoT communication anomaly detection with graph learning \cite{15}. A recent 2025 systematic review of GNN-based malicious attack detection likewise confirms the rapid expansion and diversification of this literature across network, IoT, and web-security settings \cite{16}. At the same time, evaluating HGNN-based anomaly detection models remains a significant challenge. Existing studies often use heterogeneous datasets, inconsistent evaluation metrics, and non-standardized preprocessing pipelines, making direct comparison between models difficult. Furthermore, benchmark datasets vary widely in heterogeneity, temporal scope, and label quality, which affects the reliability and reproducibility of results. Addressing this gap requires a clearer understanding of the current evaluation landscape, including what metrics and benchmarks are commonly used, how they differ, and where improvements are needed.

Despite the growing number of HGNN-based anomaly detection models, several unresolved challenges hinder their broader adoption in cybersecurity. Many approaches remain limited to static settings and struggle to capture the evolving nature of attacks and user behaviors. Data scarcity, class imbalance, and inconsistent labeling further restrict their robustness and generalizability. Moreover, interpretability and scalability remain open issues, as most models provide little explanation for detected anomalies and are computationally expensive to deploy at scale. Identifying and addressing these challenges is essential for guiding future research toward practical, interpretable, and adaptive HGNN-based solutions.

The main contributions of this survey are as follows:
\begin{itemize}
    \item \textbf{Comprehensive categorization:} We review and classify existing HGNN-based anomaly detection methods by anomaly type (node, edge, and subgraph-level) and by graph dynamics (static versus dynamic graphs).
    \item \textbf{Novel taxonomy:} We propose a unified taxonomy that captures how modeling strategies address structural and temporal heterogeneity, serving as a framework for organizing existing methods and identifying research gaps.
    \item \textbf{Application mapping:} We review four key cybersecurity domains, including insider threat detection, network intrusion detection, fraud detection in access logs, and advanced persistent threats, and analyze how HGNN models address the unique challenges and requirements of each setting.
    \item \textbf{Evaluation landscape:} We consolidate commonly used metrics and benchmark datasets, highlighting their applicability, limitations, and the need for standardized protocols.
    \item \textbf{Open challenges:} We identify limitations in current approaches and outline promising directions for future HGNN-based anomaly detection research.
\end{itemize}

The remainder of this survey is organized as follows. Section 2 reviews background on heterogeneous and dynamic graphs in cybersecurity. Section 3 presents the proposed taxonomy with a detailed discussion of node, edge, and subgraph-level approaches. Section 4 examines major application domains. Section 5 summarizes evaluation metrics and benchmark datasets. Section 6 outlines open challenges and future directions, and Section 7 concludes the survey.

\section{Background}
Graphs are a fundamental data structure for representing entities and their relationships in cybersecurity, where interactions among users, hosts, files, and network components form complex and interdependent systems. In a graph, nodes represent entities and edges represent interactions or dependencies. Traditional homogeneous graphs assume a single type of node and relation, which limits their expressiveness in real-world scenarios. In contrast, heterogeneous graphs support multiple node and edge types, capturing rich semantic information essential for modeling diverse cybersecurity behaviors. These graph structures serve as the foundation for graph-based machine learning, particularly GNNs, which have been extended to operate on heterogeneous and dynamic graphs.

\subsection{Graph Representations}
A graph is a data structure formally defined as $G = (V,E)$, where $V$ is the set of nodes and $E$ is the set of edges that represent relationships between node pairs. \cite{17}

\subsubsection{Homogeneous Graphs}
Homogeneous graphs \cite{18} are those where all nodes and edges belong to a single type and share the same features and label space. These graphs are widely used in early GNN literature and are often applied to settings such as citation networks or communication graphs, where relationships are uniform across the network. An example is an IP communication graph, where nodes represent IP addresses and edges represent undifferentiated communication links, as shown in Figure 1(a). The uniform structure simplifies both modeling and computation, making homogeneous graphs a foundational case in graph learning research.

\subsubsection{Heterogeneous Graphs}
In contrast, heterogeneous graphs \cite{18,19} involve multiple types of nodes and/or edges. A schema $S = (T_V, T_E)$ defines the set of valid node types $T_V$ and edge types $T_E$, outlining the semantic structure of the graph. Real-world cybersecurity systems often involve diverse entities (e.g., users, hosts, files) and diverse relationships (e.g., login, file access). \textcolor{blue}{These systems are naturally modeled as heterogeneous graphs that incorporate multiple node and edge types, enabling more expressive modeling of complex cyber data. Figure 1(b) illustrates this idea through a simple cyber interaction schema in which different entity categories and relation types are represented explicitly, capturing the semantic diversity that HGNNs are designed to exploit.}

\begin{figure}[htbp]
    \centering
    \includegraphics[width=0.95\columnwidth]{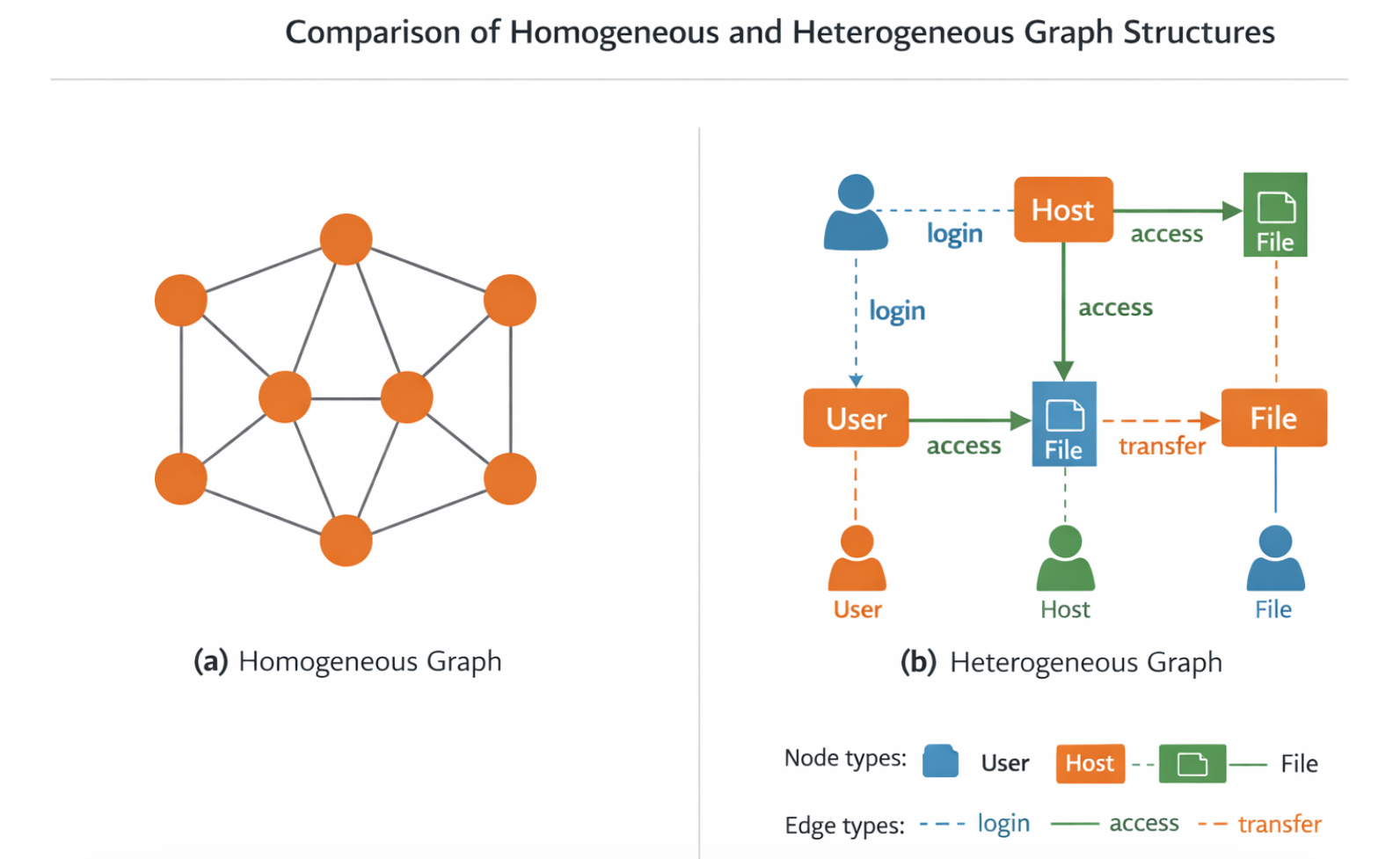}
    \caption{Comparison of homogeneous and heterogeneous graph structures. Panel (a) shows a single-type graph with uniform nodes and edges, while panel (b) shows a heterogeneous cyber graph with multiple entity and relation types.}
    \label{fig:1.1.png}
\end{figure}

\subsubsection{Graph Schema}
A graph schema defines the semantic structure of a heterogeneous graph. Formally, a schema $S = (T_V, T_E)$ consists of a set of node types $T_V$ and a set of edge types $T_E \subseteq T_V \times R \times T_V$, where $R$ is the set of relation types \cite{20}. These definitions provide a blueprint for how entities interact and are essential for designing HGNNs that can leverage semantic context. \textcolor{blue}{Schema-aware models help distinguish between normal and abnormal behaviors based on how users, systems, and files interact in multi-typed environments.}

\subsection{Graph Neural Networks: Basics}
A fundamental concept in GNNs is the node embedding, which maps each node to a vector in a continuous feature space. Nodes with similar structural roles or semantic attributes are positioned close to each other, enabling graph-structured data to be processed by standard machine learning models while preserving relational patterns. In heterogeneous graphs, embeddings must additionally encode type-specific information, ensuring that differences in node categories and relation types are reflected in the learned representations. To update these embeddings, GNNs employ a message passing mechanism. At each layer $l$, the representation of a node $v$ is updated by aggregating information from its neighbors:
  \begin{equation}
  \begin{split}
  h_v^{(l)} = \mathrm{UPDATE}^{(l)} \Bigl(
  h_v^{(l-1)}, \\
  \mathrm{AGGREGATE}^{(l)}
  \bigl(\{h_u^{(l-1)} : u \in N(v)\}\bigr)
  \Bigr)
  \end{split}
  \end{equation}

Here, $N(v)$ denotes the neighbors of $v$. The aggregation step typically uses summation, mean, or attention-weighted pooling, which ensures permutation invariance with respect to neighbor ordering, while the update step fuses the aggregated information with the node’s previous representation.

\subsection{Dynamic Graph Concepts}
Dynamic graphs extend the conventional static graph representation by incorporating temporal information into nodes, edges, and their attributes. \cite{21} While static graphs capture a single snapshot of relationships, temporal graphs describe how these relationships evolve over time, enabling the modeling of interaction sequences, evolution patterns, and time-dependent dependencies. This temporal perspective is particularly important when analyzing environments in which entity interactions are inherently time-varying.

A critical phenomenon in dynamic graph data is concept drift \cite{22}, where the statistical properties of the graph structure or associated features change over time. Such drift may arise from shifts in connectivity patterns, evolving behavioral trends, or modifications in attribute distributions. If unaccounted for, concept drift can degrade the performance of models trained on historical data, as patterns that were once indicative of normal or abnormal behavior may no longer hold. In the cybersecurity domain, dynamic graph modeling is essential for capturing the continuously changing nature of both benign and malicious activities. User behaviors, network configurations, and attack strategies are not static. They adapt in response to new technologies, security measures, and adversarial tactics. \cite{23} By representing these evolving interactions---such as user--device access patterns, file transfers, or communication flows---temporal graph analysis enables the detection of emerging anomalies and facilitates proactive threat mitigation. \cite{24} This capability goes beyond the limitations of static graph approaches, which may overlook subtle yet critical temporal signals in security-sensitive environments.

\subsection{Types of Anomalies in Graph-Structured Data}
Anomalies in graph data refer to nodes, edges, or subgraphs whose behaviors significantly deviate from patterns learned from the global topology, local neighborhood context, or temporal evolution. \cite{25} These deviations often correspond to critical events such as coordinated attacks, fraud, misconfiguration, or system failures, particularly in domains such as cybersecurity, financial systems, and industrial monitoring. In both static and dynamic graphs, anomaly detection becomes more challenging, as irregularities may emerge from evolving multi-relational semantics or shift over time. \cite{26}

Classical anomaly detection frameworks have traditionally grouped anomalies into three categories: point, contextual, and collective, depending on whether the deviation occurs independently, within a specific context, or through a set of related instances. \cite{27,28} Although conceptually comprehensive, these categories do not align directly with the structural components of graph data. Following recent surveys \cite{6,7,25}, this paper instead adopts a graph-centric taxonomy that categorizes anomalies according to the type of graph element affected. 

In this view, node-level anomalies describe individual nodes whose attributes, connectivity, or temporal behaviors deviate from normal patterns. \cite{25,29} Edge-level anomalies capture irregular or suspicious interactions between nodes such as unexpected or unauthorized links that violate relational norms. \cite{27} Subgraph-level anomalies involve groups of nodes and edges that collectively exhibit abnormal topology, density, or semantics, often reflecting coordinated or covert behavior. \cite{25,27} This taxonomy captures the full spectrum of structural irregularities observed in heterogeneous and dynamic graphs and provides a unified conceptual basis for designing and evaluating graph-based anomaly detection models. 

\subsubsection{Node Anomalies}
Node anomalies refer to individual nodes whose structural positioning, interaction patterns, or temporal behaviors deviate from the expected norms of a graph. \cite{29} These deviations may be structural (e.g., a user account with unusually low connectivity compared to peers), attribute-based (e.g., a host machine with atypical system configurations), or temporal (e.g., an employee suddenly shifting access to critical servers after months of stable behavior). In cybersecurity datasets such as CERT \cite{30} and UNSW-NB15 \cite{31}, such anomalies often correspond to insider accounts, compromised devices, or misconfigured hosts that diverge from expected interaction patterns.

To detect node anomalies in static graphs, one common approach is to examine irregularities in local connectivity or interaction density. As shown in Figure 2(a), the anomalous node appears structurally distinct from the rest of the graph because its connectivity pattern differs from that of surrounding nodes. This type of structural outlier can indicate unusual behavior, misconfiguration, or compromised activity. However, static graphs provide only a snapshot view, ignoring the temporal evolution of interactions. As a result, short-lived anomalies may disappear in aggregation, and gradual behavioral shifts may remain undetected. These limitations highlight the need for dynamic graph modeling, where temporal information is explicitly incorporated into anomaly detection.

In dynamic graphs, node-level anomalies can manifest through temporal inconsistencies, such as changes in access patterns or community affiliation. In Figure 2(b), a node that appears normal at time $t_1$ becomes anomalous at $t_2$ because its interaction pattern changes relative to the surrounding graph. This behavioral drift may indicate a role transition, lateral movement, or policy violation---common in stealthy attacks or insider threats. Capturing context-sensitive anomalies is challenging for static models, which rely on a fixed graph structure and cannot account for temporal or semantic shifts. \cite{21} In contrast, HGNN-based approaches offer enhanced capabilities for detecting complex and evolving node behaviors in heterogeneous systems by incorporating temporal reasoning, type-aware message passing, and schema-guided evolution modeling. \cite{4,19}

\begin{figure}[htbp]
    \centering
    \includegraphics[width=0.95\columnwidth]{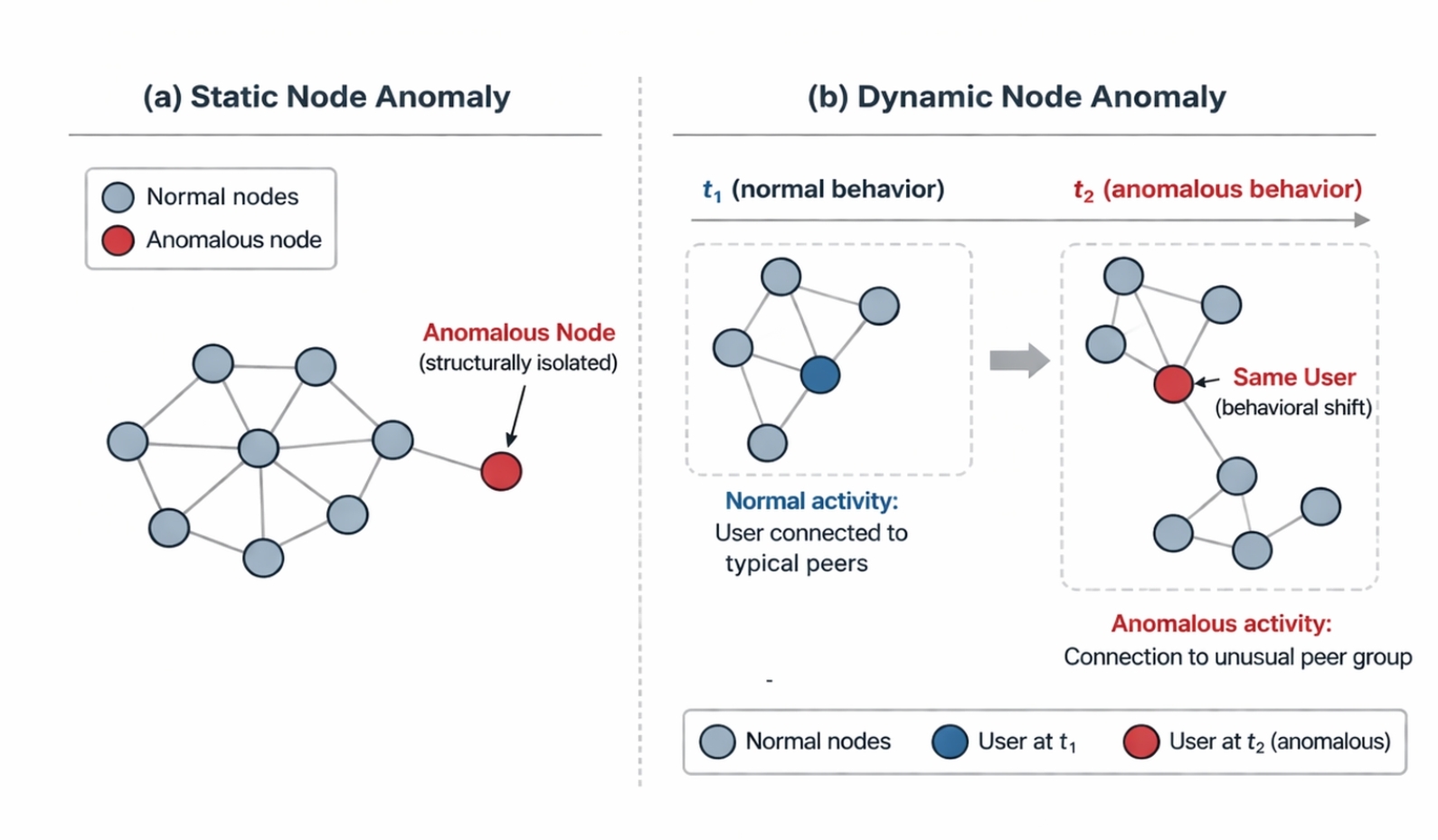}
    \caption{Node anomalies in static (a) and dynamic (b) graphs. Panel (a) illustrates a structurally atypical node within a static graph, while panel (b) shows a node whose connectivity pattern changes over time and becomes anomalous at $t_2$.}
    \label{fig:node_anomalies}
\end{figure}

\subsubsection{Edge Anomalies}
Edge anomalies describe irregular interactions between nodes that diverge from expected structural, semantic, or temporal patterns. \cite{32} These deviations are particularly challenging to detect in heterogeneous graphs, where node and edge types constrain valid relationships. In static graphs, anomalous edges often arise when a connection violates known access policies or interaction norms. As shown in Figure 3(a), one highlighted edge links an otherwise typical user-device pattern to a suspicious target, making the irregularity visible at the relation level rather than the node level alone. In cybersecurity contexts, such an anomalous edge may correspond to unauthorized access, privilege escalation, or misuse of restricted resources.

In dynamic graphs, edge anomalies are characterized by abrupt changes in connectivity over time. \cite{32} In Figure 3(b), the interaction pattern at $t_1$ appears routine, whereas at $t_2$ a new suspicious edge emerges. Such changes qualify as edge anomalies because the irregularity arises from the appearance of a new connection that deviates from historical or role-consistent patterns, rather than from the intrinsic properties of the node itself. In cybersecurity contexts, these anomalous edges may signal privilege escalation or lateral movement, where attackers establish unauthorized links to new devices or systems. Detecting these anomalies requires sensitivity to both relational semantics and temporal evolution. HGNN-based models address this by learning edge-aware representations, capturing interaction dynamics across time, and incorporating context through attention-based aggregation mechanisms. \cite{18,33}

\begin{figure}[htbp]
    \centering
    \includegraphics[width=0.95\columnwidth]{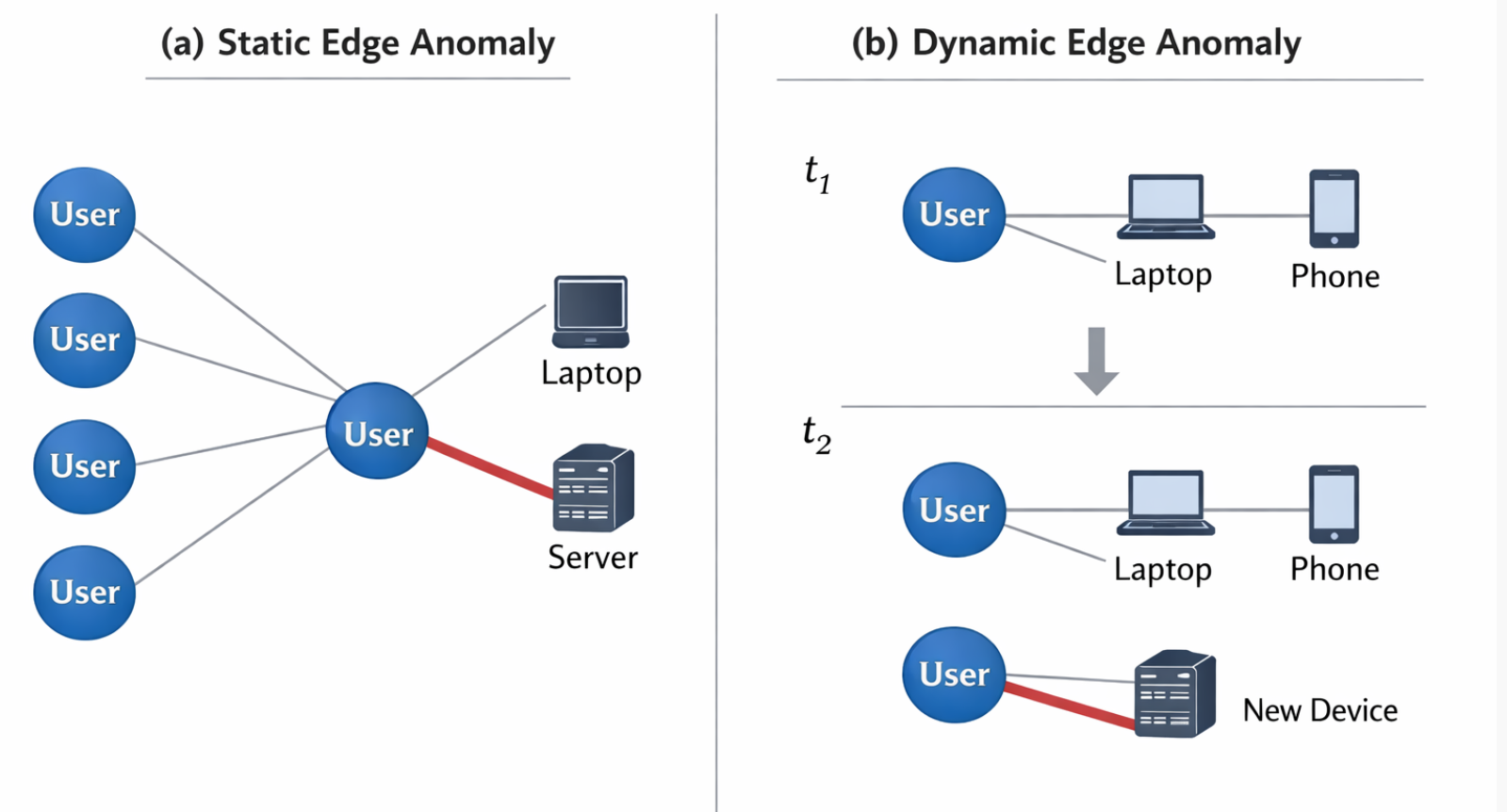}
    \caption{Edge anomalies in static (a) and dynamic (b) graphs. Panel (a) highlights an irregular edge in an otherwise normal interaction pattern, while panel (b) shows a new suspicious edge appearing over time at $t_2$.}
    \label{fig:3.1.png}
\end{figure}

\subsubsection{Subgraph Anomalies}
\begin{figure}[htbp]
    \centering
    \includegraphics[width=0.95\columnwidth]{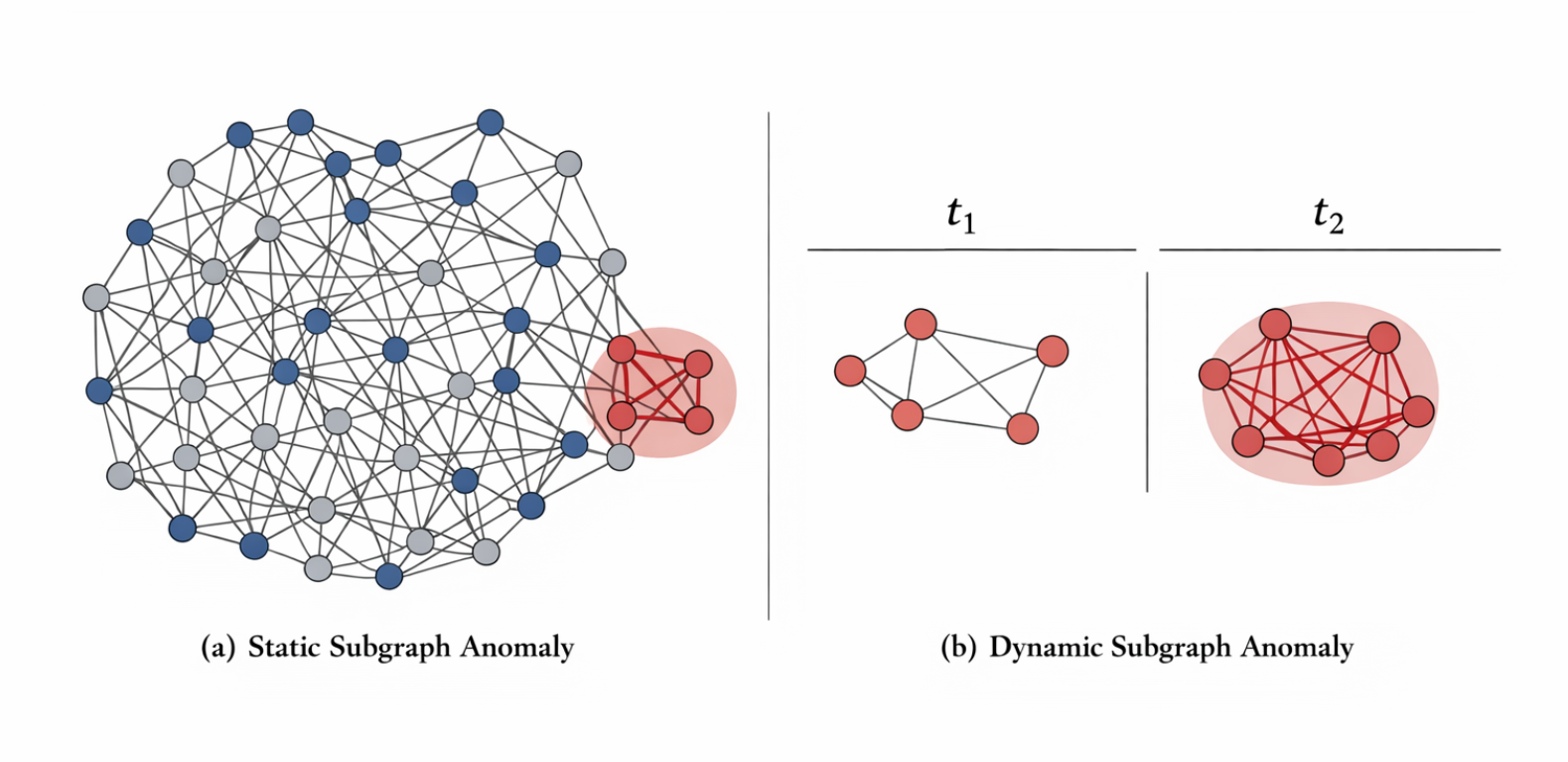}
    \caption{Subgraph anomalies in static (a) and dynamic (b) graphs. Panel (a) shows an anomalous dense local structure embedded within a larger graph, while panel (b) shows a subgraph that becomes substantially denser over time and indicates coordinated behavior at $t_2$.}
    \label{fig:4.1.png}
\end{figure}

Subgraph anomalies arise when a group of nodes and edges collectively exhibit behavior that deviates from the structural or semantic norms of the overall graph. \cite{32} Unlike node or edge anomalies, which are localized to individual elements, subgraph anomalies involve the topology, density, or temporal dynamics of a subset, often reflecting coordinated or covert behavior. In static graphs, such anomalies are typically characterized by irregular internal connectivity, motif structures, or semantic outliers. \cite{34} As shown in Figure 4(a), a compact highlighted region forms a dense local structure that stands apart from the broader graph topology. This kind of isolated dense subgraph may indicate covert collaboration, unauthorized exchanges, or coordinated malicious behavior. Recent work on subgraph anomaly detection over dynamic graphs highlights that such irregular substructures are critical indicators of coordinated attacks and evolving malicious behaviors. \cite{35} In dynamic graphs, subgraph anomalies can emerge through temporal densification or behavioral shifts. \cite{36,37} As illustrated in Figure~\ref{fig:4.1.png}(b), a relatively sparse subgraph at $t_1$ becomes much denser at $t_2$. This sudden increase in internal connectivity suggests coordinated activity that may signal collusion or synchronized access behavior. Detecting subgraph anomalies requires reasoning over higher-order structures and capturing temporal progression. HGNN-based approaches support this by enabling subgraph-level representations, modeling group semantics, and incorporating dynamic context to identify complex, evolving group behaviors. \cite{37,38,39}

Node, edge, and subgraph anomalies represent conceptually distinct categories, yet they frequently co-occur in real-world graphs. An anomalous node may simultaneously belong to an irregular substructure and participate in unexpected interactions. The perception of anomalies is often context dependent. A pattern that appears irregular in one region of a graph or at a particular time may be entirely typical in another. This complexity increases in heterogeneous graphs, where multiple node and relation types introduce rich semantic dependencies. \cite{4} Detecting such complex and evolving patterns requires models that account for structural context, semantic meaning, and temporal progression. These challenges have led to the development of graph neural frameworks that incorporate heterogeneous modeling and dynamic reasoning. \cite{1,25,26,34} \textcolor{blue}{In summary, node, edge, and subgraph anomalies represent conceptually distinct categories, yet they frequently co-occur in real-world graphs. An anomalous node may simultaneously belong to an irregular substructure and participate in unexpected interactions. Detecting such complex and evolving patterns requires models that account for structural context, semantic meaning, and temporal progression. These challenges have led to the development of graph neural frameworks that incorporate heterogeneous modeling and dynamic reasoning, which we will categorize in the next section.}

\section{Taxonomy of HGNN-Based Anomaly Detection}

To provide a unified view of existing methods, we propose a taxonomy that categorizes HGNN-based anomaly detection approaches according to the graph element they target as anomalous: node-level, edge-level, and subgraph-level. This perspective emphasizes the structural granularity of the anomaly, which fundamentally shapes both the modeling formulation and the evaluation protocol. Table~\ref{tab:taxonomy} presents the proposed taxonomy and maps representative HGNN-based anomaly detection models to each category. Unlike prior surveys that group methods by architectural choices or learning paradigms, our classification organizes the field from a task-driven perspective. This highlights common challenges and solution patterns that recur within each anomaly granularity and provides a consistent framework for comparison across methods. By structuring the discussion in this way, we enable a clearer understanding of how detection objectives influence the design of HGNN architectures, and how these designs adapt to heterogeneous and potentially dynamic graph settings. The following subsections (3.1--3.3) elaborate on each category in detail, discussing representative methods, their core mechanisms, and the specific challenges they address.

\begin{table*}[t]
\centering
\caption{Taxonomy of HGNN-based anomaly detection methods organized by anomaly granularity (node-level, edge-level, subgraph-level) and representative modeling paradigms.}
\label{tab:taxonomy}
\rowcolors{2}{gray!5}{white}
\resizebox{\textwidth}{!}{
\begin{tabular}{lp{3cm}p{3cm}p{3cm}}
\toprule
\textbf{Paradigm} & \textbf{Node-Level} & \textbf{Edge-Level} & \textbf{Subgraph-Level} \\ \midrule
Contrastive & GraphCAD \cite{40}, HeCo \cite{41} & -- & -- \\
Autoencoding & DOMINANT \cite{25}, SpecAE \cite{42} & AER-AD \cite{43}, eFraudCom \cite{44} & OCGNN \cite{45} \\
Attention & ALARM \cite{46}, GraphConsis \cite{47} & StrGNN \cite{32} & HON-GAT \cite{48} \\
Temporal & GDN \cite{49}, OCAN \cite{50}, GCAN \cite{51}, TADDY \cite{21}, HRGCN \cite{52}, XG-NID \cite{53} & AddGraph \cite{54}, DynAD \cite{55}, Bi-GCN \cite{56} & ST-GCAE \cite{57}, TGBULLY \cite{58} \\
Meta-Path & HeCo \cite{41} & -- & SubAnom \cite{35}, mHGNN \cite{59} \\
Hybrid / Motif & -- & -- & MatchGNet \cite{60}, GraphRfi \cite{61} \\
One-Class & -- & -- & OCGATL \cite{62}, OCGNN \cite{45} \\
Distillation & -- & -- & GLocalKD \cite{63} \\
Structural & -- & Hierarchical-GCN \cite{64} & MatchGNet \cite{60} \\
Adversarial & -- & AANE \cite{65} & -- \\ \bottomrule
\end{tabular}
}
\end{table*}

\begin{figure*}[t]
\centering
\includegraphics[width=\textwidth]{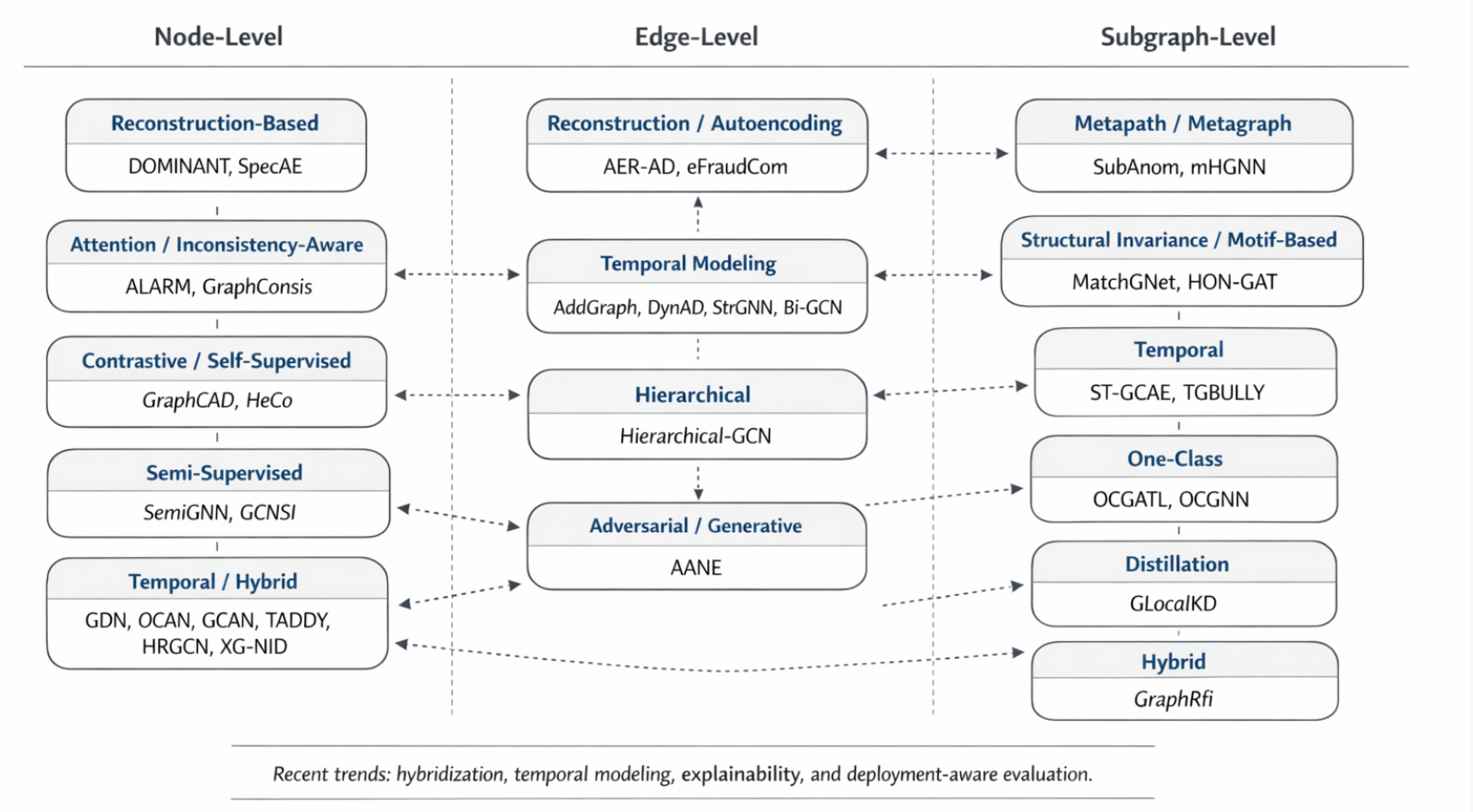}
\caption{Schematic organization of HGNN-based anomaly detection methods by anomaly granularity and dominant modeling strategy. The figure complements Table~\ref{tab:taxonomy} by visually grouping representative methods and highlighting cross-cutting trends such as temporal modeling and hybridization.}
\label{fig:schematic}
\end{figure*}

Figure~\ref{fig:schematic} provides a compact visual summary of the survey taxonomy. Unlike the tabular view in Table~\ref{tab:taxonomy}, the schematic makes it easier to see how reconstruction, temporal, structural, and hybrid modeling strategies recur across node-, edge-, and subgraph-level anomaly detection, thereby clarifying the broader organization of the literature at a glance.

\subsection{Node-Level Anomaly detection}

Node-level anomaly detection focuses on identifying individual nodes whose structural, semantic, or temporal behaviors deviate from expected patterns. \cite{66} In heterogeneous graphs, this task is particularly challenging because multiple node and edge types, role-dependent interactions, and evolving neighborhood semantics make abnormality harder to define. Over the years, a wide variety of HGNN-based models have been proposed, each grounded in different methodological principles. To systematically compare these methods, we group existing models into five categories according to their core learning strategy: reconstruction-based, attention and inconsistency-aware, contrastive learning, semi-supervised, and temporal approaches. Table~\ref{tab:node_models} presents this taxonomy, where each row corresponds to a representative model and each column specifies its supervision type, temporal characteristics, and key architectural mechanisms. This structured comparison illustrates how different modeling choices capture complementary aspects of node-level anomaly detection in heterogeneous graphs.

Reconstruction-based methods operate under the assumption that anomalous nodes are harder to reconstruct than normal ones. \textcolor{blue}{For example, \textbf{DOMINANT} \cite{25} optimizes a joint reconstruction loss:
\begin{equation}
    \mathcal{L} = (1-\alpha) \|A - \hat{A}\|_F^2 + \alpha \|X - \hat{X}\|_F^2
\end{equation}
where $A$ is the adjacency matrix and $X$ are node attributes. While effective for structural outliers, it often struggles with semantic anomalies in heterogeneous settings where relations vary significantly.} DOMINANT \cite{25} employs a dual-channel autoencoder to jointly reconstruct both adjacency structure and node attributes, while SpecAE \cite{42} extends this idea by adding spectral deconvolution and density estimation with a Gaussian Mixture Model. These approaches are straightforward and interpretable, but they often depend on strong feature homophily and can struggle when semantics are heterogeneous or evolving.

Attention and inconsistency-aware methods shift the focus from reconstruction error to semantic irregularities. ALARM \cite{46} constructs multiple GCN \cite{67} channels for different semantic neighborhoods and applies entropy-based penalties to emphasize uncertain node representations. GraphConsis \cite{47} further models inconsistencies across features, relations, and contexts, aligning these heterogeneous views to detect stealthy anomalies. These models are particularly effective in schema-rich graphs where relation diversity provides abundant semantic cues. However, they depend heavily on balanced and well-defined relation patterns, which may limit their robustness in sparse or incomplete networks. Compared with reconstruction-based methods, attention and inconsistency-aware approaches capture finer-grained relational semantics but are generally more sensitive to noise and data imbalance.

Contrastive learning provides a self-supervised alternative by aligning node embeddings across different semantic or structural views. GraphCAD \cite{40} generates semantic augmentations of local neighborhoods and applies contrastive objectives, while HeCo \cite{41} constructs dual metapath-based views and enforces consistency between them. These methods are effective in low-label scenarios and scale well to large graphs, but their performance is sensitive to the design of augmentations and view sampling strategies.

Semi-supervised approaches take advantage of limited labeled data by combining relational schema with label propagation. SemiGNN \cite{68} uses hierarchical attention over metapath-defined subgraphs for fraud detection, whereas GCNSI \cite{69} integrates GCN \cite{67} encoding with label diffusion for rumor source detection. These methods can improve accuracy under label scarcity but are highly dependent on the quality and distribution of labels, which are often uneven in practice.

Temporal models extend anomaly detection into dynamic settings by explicitly modeling node behavior over time. \textcolor{blue}{\textbf{TADDY} \cite{21} utilizes a relational transformer to encode temporal-relational interactions:
\begin{equation}
    z_i^{(t)} = \text{Transformer}(\{h_j^{(t-\tau)} : j \in N(i), \tau \in [0, T]\})
\end{equation}
By explicitly modeling the evolution of node embeddings across time steps, TADDY can detect behavioral drift that static models overlook. However, the computational complexity of transformer-based architectures remains a bottleneck for large-scale cybersecurity streaming data.} GDN \cite{49} forecasts multivariate node features and identifies anomalies as deviations from predicted values. OCAN \cite{50} couples an LSTM-based autoencoder with a GAN discriminator to refine sequence modeling, while GCAN \cite{51} fuses CNN, GRU, and GCN modules to detect propagation anomalies. TADDY \cite{21} employs a transformer-based encoder with temporal--relational attention, and recent advances such as HRGCN \cite{52} and XG-NID \cite{53} integrate hierarchical temporal cues and multi-modal input for insider threat detection. These methods capture evolving patterns effectively but introduce computational overhead and require careful temporal discretization.

\textcolor{blue}{For static enterprise graphs with well-defined schemas, attention-aware models like ALARM are preferred due to their interpretability. In contrast, for rapidly evolving network traffic where labels are unavailable, self-supervised temporal models like TADDY offer superior robustness to concept drift. Many modern frameworks are increasingly hybrid, combining temporal reasoning with contrastive objectives to alleviate label dependence while capturing dynamic patterns.}

In summary, node-level HGNN anomaly detection exhibits diverse strategies with complementary strengths. Reconstruction-based models are transparent but assume homogeneity. Attention and inconsistency-aware methods capture rich semantics but depend on diverse relational structures. Contrastive frameworks scale well without labels but rely on high-quality augmentations. Semi-supervised approaches exploit partial labels but are constrained by the quality and distribution of labels. Temporal models excel in dynamic environments but often introduce significant computational costs. Recently, research has increasingly shifted toward hybrid and self-supervised frameworks that combine multiple learning signals. Multi-view and contrastive paradigms are gaining popularity because they alleviate label dependence while enhancing representation robustness. Temporal extensions of HGNNs are also becoming more common, driven by the need to capture evolving attack behaviors and adaptive relational patterns in cybersecurity and fraud detection. Despite these advances, several challenges remain unresolved:

\begin{itemize}
    \item Limited integration of temporal reasoning, semantic consistency, and supervision within a single unified framework.
    \item Continued reliance on static and homogeneous benchmarks that fail to represent real-world heterogeneity and dynamics.
    \item Evaluation protocols that assume clean data and full observability, overlooking label noise, partial views, and concept drift.
\end{itemize}

Addressing these gaps requires end-to-end dynamic HGNNs capable of adaptive reasoning across structure, time, and semantics. Evaluating such models under realistic conditions will be essential to improve robustness, generalizability, and practical deployment in high-stakes domains such as fraud detection and cybersecurity.

\begin{table*}[t]
\centering
\caption{Comparison of HGNN models for \textbf{node-level} anomaly detection}
\label{tab:node_models}
\rowcolors{2}{gray!5}{white}
\resizebox{\textwidth}{!}{
\begin{tabular}{lllllp{4.2cm}}
\toprule
\textbf{Model} & \textbf{Learning Strategy} & \textbf{Temporal} & \textbf{Supervision} & \textbf{Key Mechanism} & \textbf{Pros/Cons} \\ \midrule
DOMINANT \cite{25} & Reconstruction & Static & Unsupervised & Dual-channel GCN autoencoder & Simple, interpretable / Assumes homophily \\
SpecAE \cite{42} & Reconstruction & Static & Unsupervised & Spectral decoder + GMM density & Captures global structure / High complexity \\
ALARM \cite{46} & Attention & Static & Unsupervised & Multi-view GCN + entropy reg. & Handles multi-view data / Sensitive to noise \\
GraphConsis \cite{47} & Inconsistency & Static & Unsupervised & Feature/Relation/Context alignment & Robust to inconsistency / Needs rich schema \\
GraphCAD \cite{40} & Contrastive & Static & Self-supervised & Semantic augmentations + contrastive MI & Scalable, no labels / Augmentation sensitive \\
HeCo \cite{41} & Contrastive & Static & Self-supervised & Dual metapath views + contrastive loss & Strong semantic views / Metapath dependent \\
SemiGNN \cite{68} & Semi-supervised & Static & Semi-supervised & Hierarchical attention over meta-paths & Uses sparse labels / Label quality dependent \\
GCNSI \cite{69} & Semi-supervised & Static & Semi-supervised & GCN encoder + label diffusion & Improves low-label performance / Task-specific \\
GDN \cite{49} & Forecasting & Dynamic & Unsupervised & Graph structure learning + feature prediction & Adaptive structure / Forecasting bias \\
OCAN \cite{50} & Forecasting & Dynamic & Unsupervised & LSTM autoencoder + GAN discriminator & Good sequential modeling / Training instability \\
GCAN \cite{51} & Forecasting & Dynamic & Unsupervised & CNN-GRU-GCN hybrid fusion & Captures propagation / Higher compute cost \\
TADDY \cite{21} & Forecasting & Dynamic & Unsupervised & Transformer encoder + temporal attention & Strong temporal modeling / Expensive \\
HRGCN \cite{52} & Forecasting & Dynamic & Unsupervised & Hierarchical relational--temporal modeling & Rich hierarchy / Complex implementation \\
XG-NID \cite{53} & Forecasting & Dynamic & Unsupervised & Transformer with multi-modal inputs & Multi-modal robustness / Resource intensive \\ \bottomrule
\end{tabular}
}
\end{table*}

\subsection{Edge-Level Anomaly Detection}

Edge-level anomaly detection focuses on identifying irregular or suspicious relationships between nodes, including the sudden creation, disappearance, or behavioral change of links. \cite{32,70} Unlike node anomalies, these irregularities are more challenging to capture because they depend not only on the characteristics of the connected nodes but also on the semantic meaning and temporal dynamics of their interactions. \cite{71} Such complexities make edge-level analysis essential for applications such as fraud detection, abnormal communication monitoring, and information flow analysis. \cite{70,72} To enable systematic comparison, existing HGNN-based models for edge-level anomaly detection are grouped into four categories based on their primary learning strategy: reconstruction and autoencoding, temporal modeling, hierarchical designs, and adversarial or generative frameworks. Table~\ref{tab:edge_models} presents this taxonomy, where each row corresponds to a representative model and each column reports its supervision type, temporal capability, and key architectural mechanism. This organization provides a clear basis for analyzing how different design philosophies address complementary aspects of edge-level anomaly detection.

Reconstruction and autoencoding methods model edge anomalies by learning embeddings that reconstruct edge presence or attributes, with deviations indicating abnormal links. AER-AD \cite{43} extends autoencoders with relational attention, explicitly scoring edges based on type-specific interactions between incident nodes. eFraudCom \cite{44} applies a graph autoencoder in transaction networks, flagging anomalous edges via reconstruction error. These models are effective in schema-rich or transaction settings but often assume that rare edges are abnormal, which risks misclassifying legitimate but infrequent interactions.

Temporal modeling approaches capture anomalies that emerge from changes in connectivity patterns over time. AddGraph \cite{54} integrates temporal attention into GCNs \cite{67}, encoding historical activation patterns in a memory module to preserve long-term dependencies. DynAD \cite{55} aggregates temporal snapshots using gated attention to adapt flexibly to streaming data. StrGNN \cite{32} combines recurrent neural networks with GCNs to jointly capture sequential dependencies and structural context, which makes it suitable for periodic or spatio-temporal anomaly patterns. Bi-GCN \cite{56} employs bidirectional message passing to detect asymmetric temporal influences, such as abnormal rumor propagation. Compared with reconstruction and autoencoding models, temporal frameworks capture evolving anomalies more effectively and maintain robustness in dynamic environments. However, they depend on assumptions about temporal granularity and often require costly retraining in non-stationary or rapidly changing settings.

Hierarchical approaches embed edges by aggregating information at multiple neighborhood scales. Hierarchical-GCN \cite{64} learns both local and global edge-centric regions, improving detection of anomalies that are invisible at the node-pair level but emerge in broader structural contexts, such as inter-community links. While this improves generalization in multiscale networks, it increases computational complexity and is sensitive to the chosen hierarchy.

Adversarial and generative frameworks learn edge embeddings by aligning them with latent priors or generating synthetic samples. AANE \cite{65} employs an adversarial autoencoder that pushes embeddings to match a learned prior, while unsupervised scoring identifies low-density edges as anomalies. These approaches offer flexibility in label-scarce settings but can struggle when benign rare interactions resemble true anomalies.

\textcolor{blue}{For edge-level detection, reconstruction methods like AER-AD are well-suited for static, schema-rich environments where anomalies manifest as rare interactions. However, in dynamic settings such as network traffic, temporal models like StrGNN and DynAD are preferred because they can capture evolving behaviors and periodic patterns. Adversarial frameworks like AANE offer strong unsupervised capabilities but may struggle with false positives in highly sparse graphs.}

In summary, edge-level anomaly detection in HGNNs reflects a variety of modeling strategies. Reconstruction and autoencoding methods are intuitive but sensitive to rarity bias. Temporal models capture evolving patterns but depend on time discretization and require higher computational overhead. Hierarchical approaches improve robustness by leveraging multi-scale contexts but add structural complexity. Adversarial frameworks enhance unsupervised performance but risk false positives in sparse graphs. Recent research in edge-level anomaly detection shows a clear shift from pure reconstruction-based models toward temporal and adversarial designs. Temporal HGNNs are gaining popularity for their ability to capture evolving interaction patterns, while adversarial and generative frameworks attract attention for improving robustness and realism in training. Early hybrid architectures that combine structural reconstruction with temporal reasoning or generative adaptation are also emerging, reflecting a growing interest in unified modeling of dynamic relational behavior.

Despite these advances, edge-level anomaly detection remains an open challenge. Current methods often address structural, temporal, or semantic irregularities separately, and few frameworks attempt to integrate these dimensions in a unified manner. Many models still assume clean relational schemas or regular temporal granularity, which limits their robustness in real-world settings where edge dynamics are noisy, asynchronous, and partially observable. Evaluations are also frequently conducted on static or synthetic benchmarks, overlooking practical factors such as evolving role semantics, bursty interactions, and context-dependent edge rarity. Future work should prioritize the development of flexible HGNN frameworks capable of jointly reasoning over structure, semantics, and time, while adapting to irregular and incomplete observations. Building richer heterogeneous benchmarks that capture realistic edge evolution will likewise be essential for improving the practical applicability of these models in domains such as fraud detection, communication networks, and cybersecurity.

\begin{table*}[t]
\centering
\caption{Comparison of HGNN models for \textbf{edge-level} anomaly detection}
\label{tab:edge_models}
\rowcolors{2}{gray!5}{white}
\resizebox{\textwidth}{!}{
\begin{tabular}{lp{3cm}llp{4.8cm}p{4.5cm}}
\toprule
\textbf{Model} & \textbf{Learning Strategy} & \textbf{Temporal} & \textbf{Supervision} & \textbf{Key Mechanism} & \textbf{Pros/Cons} \\ \midrule
AER-AD \cite{43} & Reconstruction / Autoencoding & Static & Unsupervised & Relational attention with type-specific aggregation & Good for schema-rich graphs / Rarity bias \\
eFraudCom \cite{44} & Reconstruction / Autoencoding & Static & Unsupervised & Graph autoencoder for transaction networks & Effective for fraud / Assumes rare is abnormal \\
AddGraph \cite{54} & Temporal Modeling & Dynamic & Unsupervised & Temporal GCN with attention and memory module & Long-term temporal cues / Window sensitive \\
DynAD \cite{55} & Temporal Modeling & Dynamic & Unsupervised & Gated attention over temporal snapshots & Flexible on streams / Retraining overhead \\
StrGNN \cite{32} & Temporal Modeling & Dynamic & Unsupervised & RNN-GCN fusion for spatio-temporal patterns & Strong dynamic structure / High compute \\
Bi-GCN \cite{56} & Temporal Modeling & Dynamic & Unsupervised & Bidirectional message passing for asymmetric evolution & Captures directionality / Less cyber-specific \\
Hierarchical-GCN \cite{64} & Hierarchical & Static & Unsupervised & Multi-scale neighborhood aggregation for edge-centric regions & Multi-scale context / Hierarchy sensitive \\
AANE \cite{65} & Adversarial / Generative & Static & Unsupervised & Adversarial autoencoder with latent prior alignment & Flexible unsupervised scoring / Sparse-graph false positives \\ \bottomrule
\end{tabular}
}
\end{table*}

\subsection{Subgraph-Level Anomaly Detection}

Subgraph-level anomaly detection addresses the task of identifying groups of nodes and edges that collectively exhibit abnormal behavior. \cite{73,74} Unlike node or edge-level irregularities, which are tied to individual elements, subgraph anomalies arise from the joint behavior of multiple entities that may appear normal in isolation but anomalous as a whole. \cite{75} This makes detection more challenging, particularly in heterogeneous graphs where diverse node and edge types give rise to complex relational patterns. Applications include coordinated fraud campaigns, cybersecurity monitoring, and anomaly detection in biological or chemical graphs. A wide range of HGNN-based models have been proposed for this task, but they vary considerably in their detection paradigms. To establish a coherent taxonomy, we categorize existing approaches into six groups: metapath and metagraph-based methods, structural invariance and motif-based approaches, temporal models, one-class classification, knowledge distillation, and hybrid designs. Table~\ref{tab:subgraph_models} summarizes representative models across these categories, highlighting their supervision type, temporal support, and architectural mechanisms.

Metapath and metagraph-based methods capture semantic consistency by leveraging schema-guided relational patterns. SubAnom \cite{35} constructs candidate subgraphs through metapath sampling and learns embeddings with a GNN encoder, identifying anomalies as deviations from typical semantic patterns. mHGNN \cite{59} extends this idea by aggregating higher-order semantic motifs encoded in metagraphs, which allows the model to capture multi-hop and inter-relational dependencies. While effective in networks with strong type-specific structure, these methods are heavily dependent on schema quality, and metagraph reasoning, though more expressive, introduces higher computational demands.

Structural invariance and motif-based approaches instead focus on subgraph topologies. MatchGNet \cite{60} uses hierarchical attention and subgraph matching to identify program structures that violate learned invariants, offering better generalization to unseen anomaly types. HON-GAT \cite{48} incorporates motif instances into graph attention layers, scoring subgraphs that lack expected clique or triad structures as anomalous. These approaches are robust to schema variation but depend on reliable motif definitions, and they can struggle when motif signals are weak or noisy.

Temporal models integrate subgraph evolution into the detection process. ST-GCAE \cite{57} combines spatial and temporal autoencoders to capture motif disruptions across evolving subgraphs, while TGBULLY \cite{58} models abusive behavior in social networks by applying GRU and GAT layers to evolving interaction patterns. These methods highlight the importance of temporal reasoning, particularly in detecting staged or progressive anomalies, but they increase computational cost and are sensitive to temporal discretization.

One-class classification methods are particularly useful when labeled anomalies are rare. OCGATL \cite{62} learns a compact representation boundary from normal subgraphs to isolate abnormal ones, while OCGNN \cite{45} integrates Graph Isomorphism Networks with autoencoding for molecular and transactional subgraph anomaly detection. These approaches are valuable in domains with limited supervision, though their effectiveness depends on the representativeness of normal subgraphs.

Knowledge distillation introduces a different perspective by transferring structural information across tasks. GLocalKD \cite{63} uses bi-level knowledge transfer between node-level and subgraph-level encoders, improving robustness by aligning local consistency with macro-level deviations. While effective in balancing fine- and coarse-grained information, the approach is sensitive to the alignment between teacher and student networks.

Hybrid designs combine deep embeddings with ensemble methods to improve robustness and interpretability. GraphRfi \cite{61} embeds subgraphs using GCNs \cite{67} and applies a neural random forest classifier, capturing both local and global structural abnormalities in review networks. Unlike purely embedding-based methods, GraphRfi enhances interpretability but may be less flexible across diverse graph domains.

\textcolor{blue}{For subgraph-level detection, metapath and metagraph methods like mHGNN are effective when the graph schema is well-defined and anomalies follow known semantic patterns. In contrast, structural invariance models like MatchGNet are better suited for detecting novel attack chains (e.g., APTs) where the exact semantics may vary but the underlying execution structure remains anomalous. Temporal models like ST-GCAE are essential for capturing progressive anomalies but require significant computational resources.}

In summary, subgraph-level HGNN anomaly detection demonstrates diverse modeling strategies with complementary strengths. Metapath and metagraph-based models effectively encode semantic consistency but depend on schema quality and scalability. Structural and motif-based approaches capture topological invariants but are sensitive to motif strength. Temporal methods detect evolving anomalies but at high computational cost. One-class models address label scarcity but rely on representative training data. Knowledge distillation offers cross-level robustness but requires careful alignment, while hybrid designs improve interpretability at the cost of generality.

Despite the variety of existing approaches, subgraph-level anomaly detection remains underexplored compared to node and edge-level tasks. Most models focus on either semantic consistency or structural invariance, but few frameworks attempt to integrate both dimensions in a unified manner. Temporal methods remain limited, often requiring discretized snapshots that fail to capture long-term or irregular dynamics. One-class and distillation approaches reduce label dependence but are sensitive to representation bias, while hybrid methods improve interpretability but sacrifice flexibility. Another critical gap is the scarcity of realistic benchmarks, since most evaluations rely on synthetic anomalies or domain-specific datasets that do not capture the scale and heterogeneity of real-world systems. Future work should prioritize scalable frameworks that jointly reason across structure, semantics, and time, while also addressing cross-level anomalies and evaluating under more realistic conditions.

\begin{table*}[t]
\centering
\caption{Comparison of HGNN models for \textbf{subgraph-level} anomaly detection}
\label{tab:subgraph_models}
\rowcolors{2}{gray!5}{white}
\resizebox{\textwidth}{!}{
\begin{tabular}{lp{3cm}llp{4.8cm}p{4.5cm}}
\toprule
\textbf{Model} & \textbf{Learning Strategy} & \textbf{Temporal} & \textbf{Supervision} & \textbf{Key Mechanism} & \textbf{Pros/Cons} \\ \midrule
SubAnom \cite{35} & Metapath / Subgraph Encoding & Static & Unsupervised & Candidate subgraphs via metapath sampling, GNN encoder for semantic deviation & Captures semantic patterns / Metapath dependent \\
mHGNN \cite{59} & Metagraph Reasoning & Static & Unsupervised & Aggregates higher-order motifs from metagraphs for multi-hop relations & Multi-hop reasoning / High compute cost \\
MatchGNet \cite{60} & Structural Invariance & Static & Supervised & Hierarchical attention + subgraph matcher for invariant patterns & Good for novel attack chains / Needs labels \\
HON-GAT \cite{48} & Motif-based Attention & Static & Unsupervised & Motif-instance representations integrated into attention layers & Encodes motifs / Motif quality sensitive \\
ST-GCAE \cite{57} & Temporal Modeling & Dynamic & Unsupervised & Spatial--temporal GCN autoencoder for evolving motif disruptions & Strong temporal motifs / Expensive \\
TGBULLY \cite{58} & Temporal Modeling & Dynamic & Semi-supervised & GRU + GAT layers for evolving social subgraph behavior & Captures staged behavior / Domain specific \\
OCGATL \cite{62} & One-Class Classification & Static & One-class & Learns compact boundary from normal subgraphs & Works with scarce labels / Boundary sensitivity \\
OCGNN \cite{45} & One-Class + Autoencoding & Static & One-class & GIN encoder + autoencoder for molecular/transaction subgraphs & Compact normal modeling / Benchmark limited \\
GLocalKD \cite{63} & Knowledge Distillation & Static & Supervised & Bi-level knowledge transfer between node- and subgraph-level encoders & Cross-level robustness / Teacher-student alignment needed \\
GraphRfi \cite{61} & Hybrid (Embedding + Ensemble) & Static & Supervised & GCN embeddings scored via neural random forest for review graphs & Better interpretability / Less generalizable \\ \bottomrule
\end{tabular}
}
\end{table*}

We previously discussed methods for detecting anomalous nodes, edges, and subgraphs. However, these methods share a major limitation. They primarily focus on static graphs. This static view is especially problematic for subgraph-level detection. Many key anomalies are defined by their temporal evolution. They are not just structural oddities. For instance, a dense community might form suddenly, or a group's behavior might change. Real-world networks are inherently dynamic. They constantly change in areas like cybersecurity and finance. This evolution over time often indicates an anomaly. To capture these patterns, researchers developed dynamic HGNNs. The next section will review these methods, which are designed for evolving graph data.

\section{Applications in Cybersecurity}

Cybersecurity data is inherently multi-entity, multi-relation, and evolves over time. \cite{76} Threat behaviors, whether driven by malicious insiders, external intruders, or advanced persistent actors, rarely occur in isolation. They emerge as structured patterns of interaction that span users, hosts, processes, and resources across different time periods. \cite{77} Representing these interactions as heterogeneous temporal graphs allows HGNN-based approaches to capture both the relational context and the sequential dependencies that traditional flat or tabular models often miss. This section examines four major application domains: insider threat detection, network intrusion detection, fraud detection in access logs, and advanced persistent threat (APT) or lateral movement detection. For each, we highlight how graph-based representations and reasoning make it possible to identify subtle, evolving, and context-dependent malicious behaviors in complex operational environments.

\subsection{Insider Threat Detection}

Insider threat detection focuses on identifying malicious actions perpetrated by legitimate users who exploit their access privileges to exfiltrate data, escalate access, or conduct sabotage. Unlike external attacks, insider threats often mimic normal behavior patterns, making them difficult to detect through traditional signature or rule-based systems. Graph-based modeling provides a powerful alternative by capturing the contextual relationships between users, systems, processes, and access events over time, thus enabling the discovery of subtle behavioral deviations embedded in complex relational structures.

In graph anomaly detection, insider threats are typically modeled using heterogeneous temporal graphs, where nodes represent users, machines, processes, or resources, and edges represent interaction events such as logins, file accesses, or privilege escalations. Temporal dynamics are essential, as malicious behavior often unfolds gradually or through behavioral drift. Bipartite or multipartite graphs are frequently used to separate user entities from system assets, and edge timestamps allow for modeling sequential patterns or time windows.

Several HGNN-based models discussed in Section 3 are well aligned with this task. AddGraph \cite{54} is particularly suitable for detecting access-based anomalies by modeling temporal user behavior via graph attention mechanisms. It constructs user--event graphs across discrete time windows and applies attention-based aggregation to track behavioral consistency. This allows it to flag abnormal user transitions or irregular access patterns even in unsupervised settings. GCAN \cite{51} originally developed for fake news propagation, demonstrates strong generalization to user behavior tracking by capturing sequential dependencies through its GRU-CNN-GCN hybrid architecture. In the context of insider threats, this allows for detecting suspicious propagation of commands, unusual file transfers, or unexpected account switching. OCAN \cite{50} is another representative model that leverages sequential modeling through an LSTM-based autoencoder and employs adversarial training to refine detection boundaries. The model is trained in sequences of benign user activity and uses a GAN \cite{78} discriminator to distinguish between generated and observed behavior. This approach is particularly effective for detecting insiders who slowly

evolve their behavior to avoid triggering simple statistical thresholds. OCAN has been evaluated on the CERT Insider Threat Dataset \cite{79}, a widely used benchmark consisting of simulated user activity logs from a fictitious enterprise, developed by Carnegie Mellon University. TADDY \cite{21} introduces a graph transformer framework capable of modeling long-range temporal dependencies across relation types, which is essential for capturing lateral movement or advanced persistent threats within enterprise networks. In the context of insider threats, it enables the system to detect complex temporal sequences that may include login attempts, command execution, and privilege escalations across multiple machines and sessions. Complementary enterprise-scale evidence is available in the Los Alamos National Laboratory Unified Host and Network Data Set \cite{80}, which provides authentication events, network connections, and process creation data suitable for graph-based insider threat analysis. Both datasets are high-dimensional, temporally dense, and semantically rich, making them ideal testbeds for graph-based insider threat detection research. Despite these advances, several challenges persist. Insider threat datasets suffer from extreme class imbalance, with anomalies often below one percent. Behavioral mimicry, where insiders imitate normal users, complicates detection, while gradual behavioral drift demands models that capture long-term context without overfitting to short-term noise. As a result, future research in HGNN-based insider threat detection should focus on adaptive temporal modeling, unsupervised semantic drift detection, and the fusion of structured logs with relational graph embeddings. Models that can reason over both short-term bursts and long-term deviations will be crucial to improving resilience against insider abuse.

\textcolor{blue}{Application-level results in insider threat detection remain highly dependent on the underlying release, user population, and temporal aggregation strategy, so they should not be interpreted as a direct benchmark ranking. Even so, recently reported graph-based insider threat studies show that strong performance is achievable on CERT-style benchmarks. For example, HOGPNN-ITD reports accuracy, precision, and detection rate of 0.989, 0.979, and 0.973, respectively, on CERT r4.2, while also reaching 0.972 accuracy on CERT r4.1 \cite{81}. These results suggest that temporally aware graph models can capture insider behavior effectively, but they also reinforce the need for standardized evaluation protocols before claims across studies can be compared fairly.}

\subsection{Network Intrusion Detection}
Network intrusion detection focuses on identifying unauthorized or malicious activities within computer networks, including port scanning, brute-force login attempts, data exfiltration, and protocol violations. Unlike traditional log-based methods, graph-based approaches offer a structural view of communication behavior, capturing not only the volume and frequency of connections but also the contextual and temporal relationships among communicating entities. Graph representations of network flows allow anomaly detection models to reason over communication patterns, session dynamics, and inter-device dependencies, offering improved sensitivity to stealthy or distributed attacks. Graphs used in intrusion detection are commonly constructed as IP–host graphs, host–session graphs, or temporal communication graphs, where nodes represent IP addresses, hosts, or processes, and edges denote communication sessions or flow records, often with timestamp and protocol features. Temporal windows are applied to construct evolving snapshots or streaming graphs, enabling the capture of dynamic attack behavior, such as multi-step exploits or time-distributed probes. HGNN-based anomaly detection models offer strong potential in this domain due to their ability to integrate temporal dynamics, relational heterogeneity, and context-aware representation learning. GDN, originally designed for cyber-physical systems, is well suited to network intrusion settings through its graph structure learning and multivariate time series forecasting components. In GDN, each node learns an adaptive neighborhood graph and predicts future behavior based on temporal dependencies, allowing the model to flag deviations as anomalies. Its capacity to handle multivariate node features and implicit relation learning makes it applicable to intrusion detection where relationships among traffic flows are not explicitly labeled. DynAD \cite{55} extends this dynamic modeling by explicitly learning time-evolving edge behaviors. It uses gated attention mechanisms to aggregate subgraph sequences across time and detect anomalous transitions in network topology. In intrusion detection, DynAD can model the progression of low-and-slow attacks or detect the sudden appearance of anomalous connections that deviate from historical patterns. Compared to GDN, DynAD offers finer-grained modeling of edge dynamics, which is particularly useful for capturing ephemeral connections and distributed scanning behavior. 

StrGNN \cite{32} introduces a spatio-temporal framework that fuses recurrent neural networks with GCN encoders to model both sequential activity and structural patterns. When applied to network graphs, StrGNN captures not only the topology of interactions but also periodic behaviors, such as repeated authentication attempts or scheduled transfers. Its recurrent component enables the model to maintain a memory of previous interactions, which is critical in identifying multi-phase attacks that unfold over extended periods. Bi-GCN \cite{56}, although originally proposed for rumor detection, employs bidirectional message passing to model influence propagation across graph edges. When adapted for network traffic graphs, Bi-GCN can detect asymmetric information flows indicative of command-and-control behavior, backdoors, or malicious redirection. Its bidirectional design enhances sensitivity to anomalous flow sequences that would be difficult to detect with unidirectional propagation alone. More recent graph-based IDS research reinforces this trajectory. Anomal-E \cite{9} and TS-IDS \cite{10} use self-supervised graph learning to exploit node--edge interactions under limited labels, while 2025 models such as BS-GAT \cite{13} and TE-G-SAGE \cite{14} emphasize behavior-aware graph construction, edge-aware learning, chronological evaluation, and explainability for operational settings. Complementary evidence from real-world IoT communication studies further shows that graph-based anomaly detection remains effective under low false-positive constraints and dynamic traffic conditions \cite{15}. A key strength of these models is their capacity to operate in unsupervised or weakly supervised settings, which is crucial given that network intrusion datasets often lack labeled nodes or precise attack annotations. Datasets such as UNSW-NB15 \cite{82}, CICIDS2017 \cite{83}, and CTU13 \cite{84} provide session-level or flow-level labels and have been used extensively for evaluating GNN-based intrusion detection. Graphs can be constructed by aggregating these flows over time intervals, embedding host-level attributes, and encoding session statistics (e.g., packet count, duration, protocol). However, deploying GNN-based models in real-world intrusion detection systems remains challenging. High traffic volume imposes significant computational burdens, especially for models that require neighborhood aggregation or historical memory. Furthermore, the need for streaming inference limits the feasibility of models that rely on full-batch training or multi-hop subgraph extraction. Finally, the lack of labeled ground truth for individual nodes or subgraphs hinders supervised fine-tuning and evaluation. To address these challenges, future work should focus on scalable inductive models, incremental graph construction, and semi-supervised anomaly ranking that do not require full supervision or static graphs. HGNNs capable of modeling both short-term and long-term dependencies while preserving efficient online inference will be critical for advancing the practical deployment of GNN-based network intrusion detection.

\textcolor{blue}{Although results are not directly comparable across datasets or graph-construction pipelines, published network intrusion studies do provide useful application-level evidence. On NF-BoT-IoT-v2 and NF-ToN-IoT-v2, BS-GAT reports binary-classification F1-scores of 0.9899 and 0.9790 with accuracies of 0.9900 and 0.9788, respectively \cite{13}. Under chronological evaluation on NF-UNSW-NB15-v3, TE-G-SAGE reports macro-average accuracy, precision, recall, and F1 of 0.9559, 0.4942, 0.6274, and 0.4906, outperforming a GCN baseline in recall and interpretability-oriented analysis \cite{14}. Taken together, these results support the broader conclusion of this survey: temporal and edge-aware graph models are especially valuable when intrusion detection must balance detection quality with operational interpretability.}

\subsection{Fraud Detection in Access Logs}
Fraud detection in enterprise environments involves identifying unauthorized, anomalous, or policy-violating behaviors within access control systems, enterprise resource usage, and cloud-based environments. Unlike intrusion detection which often centers on external threats, access fraud is typically conducted by legitimate users through unusual activity patterns, abuse of privileges, or coordinated circumvention of access policies. Given the complex relational structure and temporal evolution of access behaviors, graph-based modeling—particularly heterogeneous graph neural networks—has become a compelling approach for contextualizing and detecting such fraud. Access logs are naturally represented as heterogeneous graphs, where nodes can denote users, actions (e.g., login, upload), and resources (e.g., files, APIs, servers), and edges represent observed interactions over time. Graphs can be constructed from user–action–resource triplets, forming temporal session graphs with edge attributes such as timestamps, durations, or success/failure codes. These session-based graphs capture not only co-occurrence but also behavioral sequences and semantic relationships among access events, which are critical for modeling intent and consistency. 

HGNN-based models have demonstrated significant promise in this domain. SemiGNN \cite{68} introduces a semi-supervised graph attention network designed for fraud detection in financial systems. It leverages both labeled and unlabeled nodes and incorporates hierarchical attention over metapath-based subgraphs. This makes it well-suited for access fraud scenarios where only partial labels are available and fraud behaviors differ across user roles and access types. GraphConsis \cite{47} addresses fraud detection by modeling multiple sources of inconsistency in graph data—specifically, inconsistencies in features, relations, and context. This is particularly useful in enterprise logs, where fraudulent access may manifest through conflicting behavior patterns across access contexts (e.g., accessing an internal HR system from an unusual department). GraphConsis applies GCNs to multiple subgraphs views and penalizes representational divergence, effectively surfacing anomalies that violate inter-type behavioral norms. GCNSI \cite{69}, although originally proposed for rumor propagation, introduces a label propagation framework over GCN-encoded graphs. When adapted to access fraud detection, it allows for the detection of anomalous user behavior based on relational diffusion—capturing how misuse patterns may spread across user-resource interactions or departments. Its simplicity and scalability make it suitable for enterprise settings where access graphs are updated regularly. 

Common datasets for fraud detection in access logs include enterprise simulation logs, such as synthetically generated policy violation data, or real-world transaction graphs (e.g., from Alibaba or eBay platforms). While public access to corporate logs remains limited due to privacy concerns, research communities have increasingly relied on anonymized datasets and controlled simulations to benchmark graph-based fraud detectors. This domain presents unique challenges. First, behavioral heterogeneity is high—different departments, user roles, and systems generate diverse access patterns, making ``normal'' behavior difficult to define. Second, fraud itself is contextual and loosely defined; what is malicious in one role may be benign in another. Finally, label imbalance is severe: fraudulent events are rare, inconsistently labeled, and often delayed in discovery, which hampers supervised training and evaluation. Future work should address these limitations by combining HGNNs with meta-learning, domain adaptation, and semantic-aware contrastive learning, allowing models to generalize across dynamic enterprise domains and usage regimes. Semi-supervised and unsupervised approaches that incorporate access semantics, user hierarchy, and policy constraints will be key to improving robustness in fraud detection systems.

\textcolor{blue}{Public fraud-detection results are especially difficult to compare because many studies rely on proprietary interaction graphs, different fraud definitions, and different label ratios. Nevertheless, representative graph-based studies indicate clear practical gains. On public fraud benchmarks, HHLN-GNN reports improvements of 10.0\% in F1-macro, 12.5\% in AUC, and 17.3\% in G-mean on YelpChi, with smaller but still positive gains on Amazon \cite{85}. In industrial transaction settings, xFraud reports an AUC of 0.9074 on eBay-xlarge and demonstrates precision above 0.95 at operationally relevant recall levels under selected thresholds \cite{72}. These figures are illustrative rather than directly comparable, but they show that graph-based fraud detectors can provide measurable advantages when relational context is preserved.}

\subsection{Advanced Persistent Threats (APT) or Lateral Movement Detection}
Advanced Persistent Threats (APTs) refer to stealthy, multi-stage cyber intrusions that aim to maintain prolonged access within a system. Unlike one-off attacks, APTs unfold gradually through lateral movement, privilege escalation, and internal reconnaissance, often mimicking legitimate behavior to evade detection. Detecting APTs requires reasoning over long temporal sequences, diverse entity interactions, and cross-context anomalies—challenges well-suited to heterogeneous and temporal graph modeling. In practice, graph construction for APT detection typically involves host–process graphs, user–device–process graphs, or command–session graphs, where nodes represent system entities (users, machines, processes), and edges denote interactions such as command execution, session access, or process spawning. Temporal dependencies are central to this task, as APT stages (initial access, foothold, escalation, exfiltration) unfold over time and across multiple components. 

HGNN-based models address these complexities through advanced spatio-temporal and relational reasoning. MatchGNet \cite{60} introduces an invariant graph modeling framework that captures structural and semantic invariants from benign subgraphs and uses hierarchical attention to detect substructures that violate these learned templates. This makes MatchGNet particularly effective for detecting APT lateral movements, where attackers traverse unexpected host–process paths without altering node-level features significantly. GLocalKD \cite{63} enhances detection by distilling knowledge between node- and subgraph-level GNNs. In the context of APTs, this allows the model to align fine-grained local behaviors (e.g., user logins or process spawns) with high-level behavior patterns (e.g., suspicious movement through internal hosts), capturing the multi-resolution nature of threats. The joint learning strategy helps detect subtle escalations and role shifts, which may go unnoticed in models focusing solely on one level of granularity. TGBULLY \cite{58}, though originally proposed for detecting online bullying, uses a temporal GAT-GRU architecture that adapts well to capturing evolving behavior patterns in user interactions. In an enterprise setting, the same framework can be applied to session graphs where sequences of access events reflect potential staging activity for lateral movement. The model's temporal edge modeling and context attention help distinguish coordinated internal traversal from benign variability. 

The DARPA \cite{86} Transparent Computing (TC) dataset provides a comprehensive source of labeled APT behaviors, offering detailed audit logs from multiple operating systems and attack scenarios. Another valuable benchmark is the CAIDA APT simulation dataset \cite{87}, which captures network traces and command behavior from simulated APTs in a controlled environment. Despite recent progress, APT detection remains one of the most challenging applications of graph-based anomaly detection. First, APT campaigns span long temporal windows, requiring models to maintain memory over extended sequences without overfitting. Second, ground truth is sparse, making supervised training difficult and raising the need for unsupervised or weakly supervised strategies. Third, reasoning over heterogeneous relational dependencies—such as mapping user behaviors to underlying host-process interactions—requires multi-modal learning architectures that balance structural, semantic, and temporal cues. Future research should explore explainable graph transformers, online subgraph tracking, and relational motif reasoning for interpretable and real-time detection of persistent threats. Integrating domain knowledge—such as known privilege escalation paths or device roles—into HGNN architectures may further improve performance and trust in operational settings.

\textcolor{blue}{Compared with intrusion detection, quantitative evidence for APT and lateral-movement detection is still relatively sparse and often task-specific. A representative example is MatchGNet, which reports 50\% fewer false positives while maintaining zero false negatives in unknown-malware detection based on execution-behavior graphs \cite{60}. Although malware detection is not identical to enterprise APT monitoring, the result is highly relevant because it shows that invariant subgraph matching can materially reduce analyst burden while preserving detection sensitivity in attack-sequence-like settings. This also helps explain why structural and subgraph-level reasoning remains attractive for persistent-threat detection despite the limited availability of standardized APT benchmarks.}

\section{Evaluation metrics and Benchmark datasets}
Rigorous evaluation is critical for developing and comparing HGNN-based anomaly detection models. Due to the wide range of application domains and anomaly types, the community employs diverse metrics and datasets. However, evaluations are often inconsistent, and there remains a need for standardized benchmarks and reporting protocols. This section reviews commonly used evaluation metrics and benchmark datasets, structured around the taxonomy introduced in Sections 3 and 4.

\subsection{Evaluation Metrics}
The evaluation of HGNN-based anomaly detection models centers on their ability to rank true anomalies above benign instances. In practice, most methods compute anomaly scores for nodes, edges, or subgraphs, and performance is assessed using ranking-oriented metrics. The choice of metric critically affects the interpretation of model performance, especially given the prevalence of class imbalance and heterogeneous structures in real-world graphs. These metrics are used across different anomaly types: AUROC \cite{88} and AUPRC \cite{89} are most common in node-level benchmarks like ACM and DBLP. Precision@K and Recall@K are suited for streaming edge-level detection tasks like AddGraph or DynAD, where only top-ranked alerts are reviewed. NDCG@K is primarily used in subgraph-level detection settings, such as in cybersecurity or molecular domains, where anomalies carry varying degrees of severity. F1-score, while widely reported in semi-supervised node-level tasks, depends on predefined thresholds and is often used in synthetic or label-limited benchmarks. 

One of the most widely reported metrics is the Area Under the Receiver Operating Characteristic Curve (AUROC). AUROC quantifies the overall ability of the model to distinguish between anomalous and normal instances across all possible thresholds. It is computed as the area under the curve plotting the true positive rate (TPR) against the false positive rate (FPR), where:
\begin{equation}
TPR = \frac{TP}{TP + FN}, \quad FPR = \frac{FP}{FP + TN}
\end{equation}
Here, TP (true positives) is the count of correctly identified anomalies, FN (false negatives) is the count of missed anomalies, FP (false positives) is the count of incorrectly flagged normal instances, and TN (true negatives) is the count of correctly identified normal instances. Although AUROC provides a useful global ranking measure, it tends to overestimate performance in imbalanced datasets, as the false positive rate can remain deceptively low when the normal class dominates. To counter this, the Area Under the Precision-Recall Curve (AUPRC) is often recommended, particularly in settings with severe class imbalance such as fraud detection and insider threat scenarios. The precision (P) and recall (R) are defined as:
\begin{equation}
Precision = \frac{TP}{TP + FP}, \quad Recall = \frac{TP}{TP + FN}
\end{equation}
AUPRC emphasizes performance on the minority class (anomalies), making it a more informative measure than AUROC in many HGNN use cases. 

For practical deployment scenarios—where analysts typically investigate only the top-ranked results—metrics such as Precision@K and Recall@K provide actionable insight. Precision@K measures the fraction of true anomalies within the top K ranked predictions:
\begin{equation}
Precision@K = \frac{\text{Number of True Anomalies in Top } K}{K}
\end{equation}
Similarly, Recall@K measures the proportion of total anomalies that appear within the top K results. In subgraph-level detection tasks, where anomalies can vary in severity (e.g., minor vs. severe breaches), models like MatchGNet and GLocalKD utilize the Normalized Discounted Cumulative Gain (NDCG) to incorporate graded relevance. NDCG@K is defined as:
\begin{equation}
NDCG@K = \frac{1}{IDCG@K} \sum_{i=1}^{K} \frac{2^{rel_i} - 1}{\log_2(i + 1)}
\end{equation}
where $rel_i$ represents the relevance score (or severity) of the anomaly at the $i$-th ranked position, and IDCG@K is the ideal (best possible) DCG value for normalization. In semi-supervised models where partial ground truth is available, F1-score is frequently reported. It combines precision and recall into a single metric:
\begin{equation}
F1 = 2 \times \frac{Precision \times Recall}{Precision + Recall}
\end{equation}
Although intuitive, the F1-score depends on a predefined threshold, which can be difficult to set reliably in unsupervised anomaly detection tasks unless threshold calibration is handled systematically. 

Recent studies have underscored common pitfalls in metric selection and reporting. These include threshold sensitivity, potential inflation of AUROC in highly imbalanced datasets, and a lack of standardized operational evaluation protocols. To mitigate these challenges, current best practices emphasize reporting multiple complementary metrics, ensuring clarity around threshold selection, and aligning evaluations with the intended deployment context. Table 5 summarizes the core evaluation metrics discussed above, including their mathematical formulations and recommended usage contexts within HGNN-based anomaly detection.

\begin{table*}[t]
\centering
\caption{Core Evaluation Metrics in HGNN-Based Anomaly Detection}
\label{tab:metrics}
\rowcolors{2}{gray!5}{white}
\resizebox{\textwidth}{!}{
\begin{tabular}{lp{5cm}p{6cm}}
\toprule
\textbf{Metric} & \textbf{Formula / Description} & \textbf{Application Context} \\ \midrule
AUROC & AUROC $= \int_0^1 TPR(FPR^{-1}(x)) dx$ & Node-level detection (e.g., DOMINANT, HeCo); standard for global ranking. \\
AUPRC & AUPRC $= \int_0^1 P(R^{-1}(x)) dx$ & Node-level in imbalanced datasets (GraphConsis, OCAN); robust to false positives. \\
Precision@K & Precision@K = $\frac{\text{True Anomalies in Top K}}{K}$ & Edge-level, real-time detection (AddGraph, DynAD); used in limited analyst settings. \\
Recall@K & Recall@K = $\frac{\text{True Anomalies in Top K}}{\text{Total Anomalies}}$ & Edge- or subgraph-level (GCAN, DynAD); ranks based on high-sensitivity detection. \\
F1-score & $F1 = 2 \times \frac{\text{Precision} \times \text{Recall}}{\text{Precision} + \text{Recall}}$ & Semi-supervised node detection (SemiGNN, GCNSI); sensitive to thresholding. \\
NDCG@K & $NDCG@K = \frac{1}{IDCG_K} \sum_{i=1}^K \frac{2^{rel_i} - 1}{\log_2(i+1)}$ & Subgraph-level (MatchGNet, GLocalKD); ranks based on graded threat relevance. \\ \bottomrule
\end{tabular}
}
\end{table*}
Given the wide range of anomaly structures and task formulations, no single metric provides a complete picture. As such, recent studies increasingly advocate for multi-metric evaluation strategies that combine global ranking metrics (e.g., AUROC) with task-oriented metrics (e.g., Precision@K or NDCG), especially when evaluating models under real-world operational constraints. \textcolor{blue}{In particular, for streaming cybersecurity systems, \textbf{Precision@K} is often the most critical metric as it directly correlates with the false alert fatigue experienced by security analysts. A high AUROC may be misleading if the top-ranked results are dominated by false positives, rendering the system unusable in practice.}

\subsection{Benchmark Datasets}
The choice of benchmark datasets strongly influences both model design and evaluation outcomes. In the context of heterogeneous graphs, benchmark datasets must capture multi-typed node and edge structures, temporal evolution, and realistic anomaly patterns. Existing benchmarks vary widely in annotation quality, graph construction assumptions, and temporal richness, often posing challenges for consistent and reproducible evaluation. This subsection reviews commonly used datasets across node-level, edge-level, and subgraph-level anomaly detection tasks, detailing how graphs are constructed from raw data and examining critical factors such as heterogeneity, temporal scope, and label quality. By highlighting dataset characteristics and limitations, we aim to clarify where current evaluation practice remains weak and what properties future benchmarks should provide.

\subsubsection{Node-Level Anomalies}
A significant proportion of prior work utilizes semi-synthetic citation networks such as Cora \cite{90}, Citeseer \cite{90}, and Pubmed \cite{90}. These datasets represent homogeneous graphs where node features and links are artificially perturbed to simulate anomalies, for instance, through attribute noise injection or random edge rewiring. Models such as DOMINANT \cite{25} and SpecAE \cite{42} adopt these benchmarks in semi-supervised settings. While widely adopted, these datasets offer limited realism for heterogeneous or dynamic scenarios, and the anomalies they model are often simplistic. To address this, heterogeneous datasets such as Amazon \cite{91}, Yelp \cite{92}, and DBLP \cite{93} have been employed in more recent models like HeCo \cite{41} and GraphCAD \cite{40}, which construct metapath-based user–item interaction graphs. These benchmarks better reflect real-world heterogeneity but present challenges such as schema ambiguity, unclear ground truth labels, and manual metapath specification. Additionally, datasets like eBay extend node-level modeling into transactional domains, capturing user behaviors in online marketplaces. While rich in relational semantics, eBay data is often incomplete or access-restricted, making reproducibility and evaluation difficult.

\subsubsection{Edge-Level Anomalies}
For edge-level detection tasks, CICIDS2017 \cite{83} and CTU-13 \cite{84} are widely used. These datasets consist of temporal graphs modeling network flows between IP addresses and services, with ground-truth labels for attacks such as port scans, brute-force intrusions, and botnet activity. Models like DynAD \cite{55}, StrGNN \cite{32}, and SedanSpot \cite{94} utilize these benchmarks to identify anomalous interactions. However, they are characterized by high class imbalance and sparsely labeled attack events, which necessitate careful evaluation using metrics like AUPRC and Recall@K. The eBay transaction graph \cite{95} is also used in edge-level anomaly detection, particularly in financial fraud scenarios. Here, edges represent transactions, and anomalies correspond to fraudulent or suspicious interactions. Due to limited annotation, this dataset is often handled through semi-supervised or weakly supervised learning frameworks, as in models like eFraudCom.

\subsubsection{Subgraph-Level Anomalies}
Subgraph-level anomaly detection remains less developed, with relatively few public benchmarks. Notable examples include the DARPA \cite{86} Transparent Computing (TC) dataset, which captures host–process and user–device–process interactions over time in enterprise networks. This dataset is used in models like MatchGNet and GLocalKD to detect complex multistep attack chains. Similarly, CAIDA APT \cite{87} provides network traces involving advanced persistent threats (APTs), enabling evaluation of models such as TGBULLY and ST-GCAE that focus on long-term temporal anomalies and subgraph evolution. The Open Graph Benchmark (OGB) Series \cite{96} is another important resource. While primarily used for node or graph-level classification, its segmentation-based subgraph support has been adapted in models like OCGNN \cite{45} and HON-GAT. However, OGB benchmarks often lack detailed subgraph-level anomaly labels, limiting their direct applicability.

\subsubsection{Cross-Cutting Issues}
Across all datasets, a recurring limitation is the lack of standardization in graph construction protocols. Studies often apply different preprocessing techniques, metapath definitions, time windowing methods, edge aggregation strategies, or feature engineering pipelines, which can significantly impact the resulting graph topology and feature distributions and thereby hinder reproducibility and fair comparison. Additionally, while many benchmarks are static snapshots, real-world anomaly detection increasingly requires datasets that support streaming, online detection, and temporal drift. Recent efforts like SedanSpot, evaluated on CICIDS2017 \cite{83}, attempt to address this by simulating streaming edge environments, but broader community adoption of dynamic benchmarking remains limited.

In summary, while existing benchmarks have laid the foundation for HGNN-based anomaly detection, there remains a critical need for diverse, standardized, and temporally aware datasets, particularly in heterogeneous or multi-phase attack scenarios. Table~\ref{tab:benchmarks} summarizes the key characteristics of each benchmark dataset discussed in this section. As shown in Table~\ref{tab:benchmarks}, while several datasets exist for evaluating HGNNs, there is a clear gap in high-heterogeneity, dynamic benchmarks tailored for advanced threats like APTs. Addressing this gap will be crucial for advancing the field and ensuring that models are evaluated under realistic, operational conditions.

\begin{table*}[t]
\centering
\caption{Summary of Benchmark Datasets for HGNN-Based Anomaly Detection}
\label{tab:benchmarks}
\rowcolors{2}{gray!5}{white}
\resizebox{\textwidth}{!}{
\begin{tabular}{lp{2.8cm}p{4.2cm}llp{3cm}p{3.2cm}}
\toprule
\textbf{Dataset} & \textbf{Anomaly Type} & \textbf{Graph Construction} & \textbf{Heterogeneity} & \textbf{Temporal} & \textbf{Example Models} & \textbf{Deployment Context} \\ \midrule
Cora, Citeseer, Pubmed \cite{90} & Node-Level (Synthetic) & Citation graphs with attribute/edge perturbations & Homogeneous (synthetic in HGNNs) & Static & DOMINANT, SpecAE & Academic / Baseline \\
Amazon \cite{91}, Yelp \cite{92}, DBLP \cite{93} & Node-Level & User--item metapath graphs & Heterogeneous & Static & HeCo, GraphCAD & E-commerce / Review Fraud \\
CICIDS2017 \cite{83} & Edge-Level & Network flow graphs between IPs; temporal snapshots & Semi-heterogeneous (IP, protocol) & Temporal & SedanSpot, DynAD & Network Intrusion \\
CTU-13 \cite{84} & Edge-Level & Botnet flow graphs & Semi-heterogeneous & Temporal & DynAD, StrGNN & Botnet Detection \\
eBay \cite{95} & Edge-Level & Transaction graphs (users, transactions) & Heterogeneous & Static & eFraudCom & Financial Fraud \\
CAIDA APT \cite{87} & Subgraph-Level & Network traces capturing advanced persistent threats & Heterogeneous & Temporal & TGBULLY, ST-GCAE & APT / Lateral Movement \\
OGB Series \cite{96} & Subgraph-Level & Citation/biological graphs; subgraphs via segmentation & Varies & Static & OCGNN, HON-GAT & General Graph Learning \\ \bottomrule
\end{tabular}
}
\end{table*}

\subsubsection{\textcolor{blue}{Targeted Empirical Case Study}}
To complement the survey discussion with concrete evidence, we conducted a targeted empirical case study using public-code implementations of representative graph anomaly detection models. Rather than attempting an exhaustive benchmark across all methods reviewed in this survey, we selected three reproducible models that represent distinct design philosophies: a static reconstruction-based baseline (DOMINANT), a dynamic transformer-based model (TADDY), and a structural-temporal graph model (StrGNN). All three models were executed under a unified local preprocessing pipeline that converted the public UNSW-NB15 release into model-specific inputs.

Because the publicly accessible UNSW-NB15 release does not expose raw host identities in a form directly usable for host-to-host graph construction, we generated a documented behavioral-profile proxy graph from the available flow attributes. Consequently, the following results should be interpreted as a reproducible case study under a common graph-construction protocol rather than as directly comparable benchmark numbers from the original papers. Even with this limitation, the experiment is useful because it illustrates how representative static, temporal, and structural-temporal methods behave under the same practical preprocessing assumptions.

\begin{table*}[t]
\centering
\caption{\textcolor{blue}{Targeted empirical case study under a unified public-code pipeline. Results are reported on the normalized public UNSW-NB15 release and should be interpreted as case-study outputs under a reproducible local graph-construction protocol.}}
\label{tab:case_study}
\rowcolors{2}{gray!5}{white}
\begin{tabular}{llll}
\toprule
\textbf{Model} & \textbf{Paradigm} & \textbf{Setting} & \textbf{Observed Result} \\ \midrule
DOMINANT \cite{25} & Static reconstruction & 20 epochs & AUROC $\approx 0.365$ \\
TADDY \cite{21} & Dynamic transformer & 5 epochs & Total AUC $\approx 0.639$ \\
StrGNN \cite{32} & Structural-temporal GNN & 5-epoch smoke test & AUROC $\approx 0.733$, AP $\approx 0.730$ \\ \bottomrule
\end{tabular}
\end{table*}

The results in Table~\ref{tab:case_study} show that executable reproduction is feasible, but performance is sensitive to graph construction choices. On the prepared UNSW-NB15 proxy graph, DOMINANT achieved a smoke-test AUROC of approximately 0.365, TADDY achieved a total AUC of approximately 0.639, and StrGNN achieved an AUROC of approximately 0.733 with average precision of approximately 0.730. Within this unified setup, the structural-temporal model produced the strongest anomaly-separation behavior, while the dynamic transformer remained clearly stronger than the static reconstruction baseline. These observations are directionally consistent with the broader survey argument that temporal and structure-aware models are better suited to evolving cybersecurity anomalies than purely static methods.

\begin{figure*}[t]
\centering
\includegraphics[width=\textwidth]{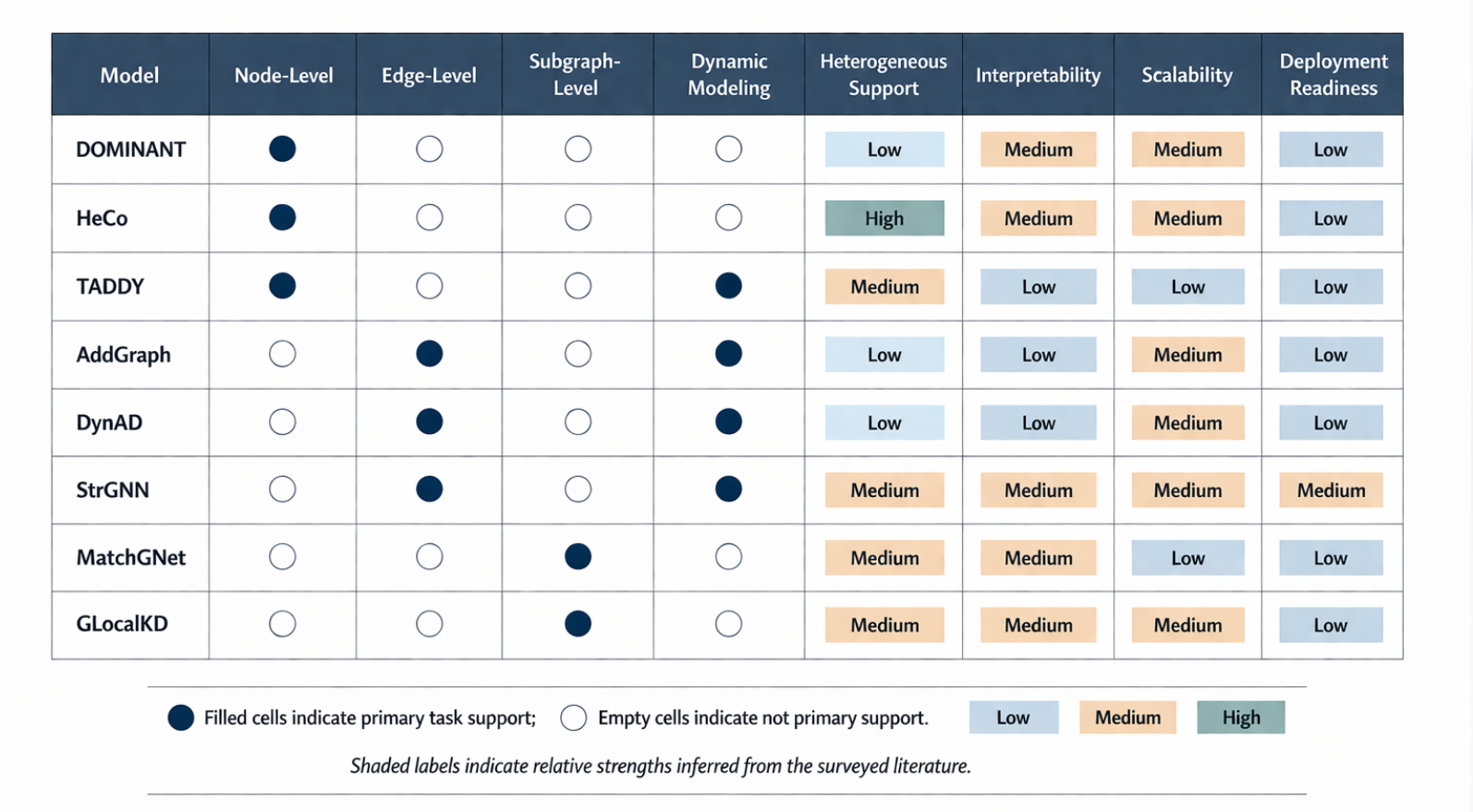}
\caption{Visual comparison of representative HGNN-based anomaly detection models across anomaly granularity, temporal capability, heterogeneous support, interpretability, scalability, and deployment readiness. Filled cells indicate primary task support, while shaded labels indicate relative strengths inferred from the surveyed literature.}
\label{fig:comparison}
\end{figure*}

Figure~\ref{fig:comparison} complements the empirical case study by providing a compact cross-model comparison that is difficult to convey through text alone. In particular, it highlights the trade-offs between dynamic modeling strength, heterogeneous support, interpretability, scalability, and deployment readiness, helping readers quickly identify which representative models are better aligned with different cybersecurity settings.

\subsubsection{Concrete Recommendations for Evaluation}
\textcolor{blue}{
To address the limitations identified in current datasets and move toward operationally relevant benchmarking, we propose the following recommendations:
\begin{enumerate}
    \item \textbf{Standardized Temporal Splits:} Use chronological splits (e.g., 70/10/20\% by time) instead of random shuffling to prevent ``looking into the future'' and to evaluate model robustness against concept drift.
    \item \textbf{Imbalance-Aware Metrics:} Report both AUROC and AUPRC. Given that anomalies often represent less than 1\% of data, AUPRC provides a more realistic measure of a model's ability to isolate rare threats without excessive false positives.
    \item \textbf{Analyst-Centric Metrics (Precision@$K$):} Evaluate models using Precision@$K$, where $K$ is calibrated to the typical daily alert capacity of a Security Operations Center (SOC) (e.g., $K = 50$ or $100$). This directly measures the alert-fatigue impact on human analysts.
    \item \textbf{Heterogeneity Reporting:} Explicitly report the number of node and edge types utilized. Models that achieve high performance by collapsing a heterogeneous graph into a homogeneous one should be critically evaluated for their loss of semantic nuance.
\end{enumerate}
}

\subsubsection{Practical Deployment Considerations}
\textcolor{blue}{
Beyond traditional metrics, the transition from research to production requires addressing several deployment constraints:
\begin{enumerate}
    \item \textbf{Latency Budgets:} Benchmark end-to-end inference latency under realistic streaming or batched conditions, since cybersecurity alerts lose value when they arrive too late for operational response.
    \item \textbf{Memory Footprint:} Report GPU/CPU memory consumption during both training and inference, especially for temporal and transformer-based HGNNs intended for enterprise-scale graphs.
    \item \textbf{Explainability at Triage Time:} Measure whether the model can surface supporting entities, relations, or substructures quickly enough for analyst validation, rather than treating explainability as an offline-only feature.
    \item \textbf{Drift Robustness:} Evaluate how performance degrades over time without retraining, and report the refresh cadence needed to maintain acceptable performance in changing environments.
\end{enumerate}
}

\section{Open Challenges and Future Directions}
Despite notable progress in heterogeneous graph neural networks (HGNNs) for anomaly detection, several obstacles remain before such methods can be reliably deployed in real-world cybersecurity systems \cite{3,18,75}. These challenges span modeling limitations, data availability, evaluation practices, and deployment feasibility. Recent 2025 literature continues to highlight the same pressure points, especially explainability, realistic graph construction, chronology-aware evaluation, and robustness under operational traffic conditions \cite{13,14,15,16}. Addressing them is essential for advancing both methodological rigor and operational readiness.

\begin{figure*}[t]
    \centering
    \includegraphics[width=\textwidth]{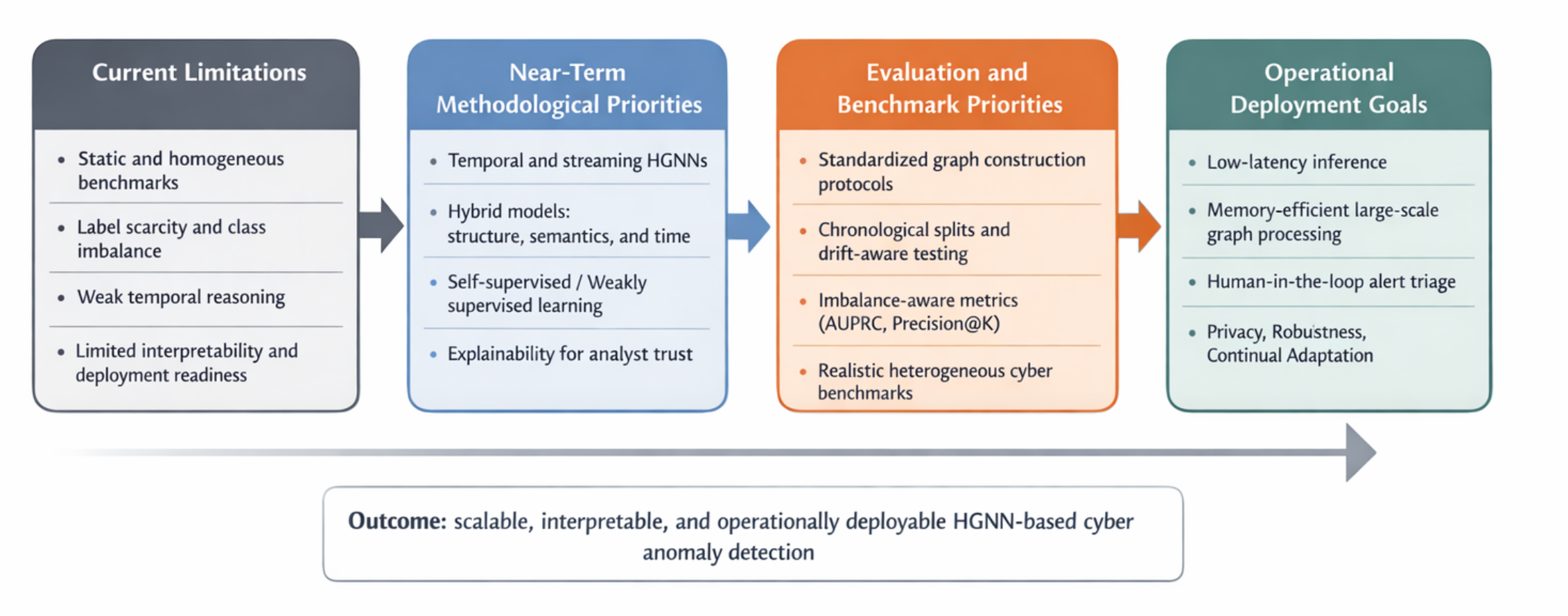}
    \caption{Roadmap for advancing HGNN-based anomaly detection in cybersecurity. The figure summarizes current limitations, near-term methodological priorities, evaluation and benchmark needs, and operational deployment goals required to move HGNN-based cyber anomaly detection toward practical use.}
    \label{fig:roadmap}
\end{figure*}

Figure~\ref{fig:roadmap} provides a structured summary of the major research directions highlighted in this section. It organizes the field's next steps from current limitations through methodological and evaluation priorities to deployment goals, emphasizing that progress in HGNN-based cybersecurity anomaly detection depends not only on stronger models, but also on realistic benchmarks, reproducible evaluation, and operationally grounded system design.

\textcolor{blue}{Despite strong reported results, many current HGNN-based anomaly detection methods still struggle in practice for five recurring reasons. First, their performance is highly sensitive to graph construction choices, including node and edge definitions, temporal windowing, and schema design, which are rarely standardized across studies. Second, many models assume relatively clean and stationary relational patterns, whereas real cybersecurity environments are noisy, incomplete, and subject to concept drift. Third, benchmark gains often depend on offline training and broad graph access, making them difficult to transfer to low-latency streaming settings. Fourth, anomaly scores are frequently insufficiently interpretable for analyst triage, limiting operational trust even when ranking performance is high. Fifth, many published evaluations emphasize detection performance in isolation, without accounting for memory footprint, alert fatigue, retraining cost, or integration into existing SOC workflows. As a result, the gap between benchmark success and deployment readiness remains a central weakness of the current literature.}

\textcolor{blue}{Table~\ref{tab:practice_gaps} summarizes these recurring failure sources, how they appear in the current literature, and why they continue to limit practical deployment in cybersecurity settings.}

\begin{table*}[t]
\centering
\caption{\textcolor{blue}{Why current HGNN-based anomaly detection methods still struggle in practice}}
\label{tab:practice_gaps}
\rowcolors{2}{gray!5}{white}
\resizebox{\textwidth}{!}{
\begin{tabular}{p{3.3cm}p{5.6cm}p{5.6cm}}
\toprule
\textbf{Failure Source} & \textbf{Manifestation in Current Literature} & \textbf{Operational Consequence} \\ \midrule
Graph construction instability & Results change substantially with different node/edge definitions, temporal windows, and heterogeneous schemas; these choices are often underreported. & Weak cross-paper comparability and poor transfer from published benchmarks to real deployments. \\
Distribution shift and concept drift & Models are commonly trained on static or quasi-static data, while operational cyber environments evolve continuously. & Rapid performance degradation, retraining burden, and missed emerging threats. \\
Limited interpretability & Many methods output anomaly scores without clear supporting entities, relations, or subgraphs for analyst validation. & Low analyst trust, slower triage, and higher false-alert fatigue. \\
Computational and memory overhead & Temporal, transformer-based, and subgraph-centric methods often require expensive aggregation, history storage, or offline processing. & Difficulty meeting real-time latency budgets and scaling to enterprise graphs. \\
Weak benchmark realism & Common datasets are synthetic, simplified, or inconsistently preprocessed, with limited heterogeneity or sparse temporal labels. & Inflated experimental results that do not reliably predict field performance. \\
Evaluation misalignment & Studies frequently optimize AUROC or similar metrics without measuring deployment costs, alert capacity, or maintenance effort. & Models may look strong offline yet remain impractical for SOC integration. \\ \bottomrule
\end{tabular}
}
\end{table*}

\textcolor{blue}{Taken together, these gaps explain why strong benchmark performance does not automatically translate into operational usefulness. They also motivate the more specific modeling, data, and evaluation challenges discussed in the following subsections.}

\subsection{Modeling Challenges}
Cybersecurity graphs are complex and constantly evolving, creating several open challenges for HGNN-based anomaly detection \cite{17,66,64}. A key issue is temporal dynamics: many models rely on static graph snapshots and fail to capture long-term dependencies that are critical for identifying staged attacks or gradual privilege escalation \cite{29,55}. Another challenge lies in heterogeneity and modality, since real systems combine multiple entity types and often include logs, textual reports, or event traces that current approaches rarely exploit \cite{22}.

A further challenge concerns interpretability. Anomaly scores generated by HGNNs are often opaque, which hinders analysts from understanding or trusting model outputs in high-stakes cybersecurity environments. In operational settings, explainability is not merely desirable but essential for incident investigation, compliance auditing, and human--AI collaboration. Although attention mechanisms and feature attribution have been explored to provide interpretive insights \cite{25,39}, these methods typically offer coarse-grained explanations and are difficult to extend to heterogeneous graphs, where multiple relation types and semantics coexist. Future research should focus on integrating fine-grained interpretability within HGNN architectures, enabling anomaly reasoning that is traceable to specific entities, relations, and causal contexts.

Future research should also emphasize hybrid modeling architectures that combine temporal reasoning, multimodal fusion, and interpretable learning. Promising directions include graph transformers for cross-type dependency modeling \cite{25}, neuro-symbolic reasoning for explainable anomaly inference, and causal or contrastive frameworks that can disentangle spurious correlations from meaningful behavioral deviations \cite{39}. Advancing these directions will require balancing transparency, scalability, and robustness within unified, end-to-end HGNN architectures.

\subsection{Data Challenges}
Data limitations remain one of the most significant barriers to advancing HGNN-based anomaly detection \cite{4,69}. A major challenge is label scarcity and imbalance: malicious events are rare and costly to annotate, while benign activities dominate most datasets \cite{26,36}. This imbalance makes models prone to bias and reduces their ability to detect rare but high-impact anomalies. Semi-supervised, self-supervised, and weakly supervised learning strategies show promise, yet they require careful adaptation to heterogeneous and temporal graph settings \cite{67}.

Another persistent issue is benchmark realism. Many publicly available datasets are synthetic, outdated, or lack the structural and temporal complexity observed in operational systems \cite{33}. As a result, models that perform well on controlled datasets often fail to generalize in production environments. Closing this simulation-to-real gap will require large-scale, richly annotated, and temporally consistent benchmarks that capture noise, incomplete observability, and evolving adversarial tactics \cite{49}. Collaboration between academia and industry could facilitate the creation of shared, privacy-preserving cybersecurity graph datasets. Future directions include federated graph learning for secure data sharing \cite{51}, synthetic data generation to augment rare attack cases, and transfer learning pipelines to adapt HGNNs across different environments while respecting privacy and confidentiality constraints \cite{52,57}.

\subsection{Evaluation Challenges}
Current studies employ highly varied evaluation metrics and graph construction protocols, which complicates cross-paper comparison and limits reproducibility \cite{5,27}. Without standardized practices, it becomes difficult to measure progress or validate new methods consistently. Establishing community-agreed benchmarks with unified metrics (e.g., AUROC, Precision@$K$) and consistent preprocessing pipelines would enable fairer comparison and stronger claims of improvement \cite{6,23}.

Beyond metric unification, the field needs reproducible and transparent evaluation frameworks \cite{24}. Public leaderboards, standardized data splits, and open-source codebases can encourage more rigorous benchmarking and reduce fragmentation. Another pressing need is for robust evaluation under realistic conditions, such as noisy, partially observable, or evolving graphs \cite{41,58}. These efforts will help ensure that reported improvements reflect genuine robustness and operational value rather than overfitting to idealized datasets.

\subsection{Practical Deployment Challenges}
Deploying HGNN-based anomaly detection in real environments raises critical challenges of scalability, adaptability, and operational reliability \cite{28}. Cybersecurity graphs often contain millions of nodes and edges that evolve in near real time, yet many existing HGNN architectures remain too computationally expensive for production use \cite{30,43}. Real-time decision latency is another major constraint: anomaly detectors must process streaming updates and generate alerts within strict time budgets to support timely incident response. Achieving this balance between inference speed and detection accuracy requires efficient message-passing schemes, streaming graph processing, and resource-aware model compression \cite{71}.

Beyond latency and scale, deployment also introduces security and privacy concerns. HGNNs may themselves become targets of adversarial manipulation, including model poisoning or evasion attacks that exploit their learned representations \cite{32}. Moreover, training on sensitive logs or network data raises privacy and compliance risks, especially when graph embeddings inadvertently encode identifiable or confidential information \cite{38}. Mitigating these risks calls for privacy-preserving learning techniques such as federated or encrypted graph learning, as well as robust model auditing, explainability dashboards, and access-control mechanisms.

Operational environments further demand reliability, interpretability, and maintainability \cite{68}. Threat behaviors evolve continuously, and static models can quickly degrade as attack patterns shift \cite{54}. Online and continual learning techniques are required to adapt parameters as new data arrive without catastrophic forgetting \cite{56,45}. Equally important is interpretability at the deployment stage: transparent anomaly explanations allow analysts to validate alerts efficiently, reduce false positives, and improve overall response speed. Designing human-in-the-loop pipelines that integrate interpretability with adaptive learning will be essential to ensure operational trust and long-term resilience.

Future progress will depend on the development of lightweight, secure, and adaptive HGNN frameworks that can meet real-time constraints while preserving privacy and interpretability. Integrating scalable inference, adversarial robustness, and continual learning into unified operational pipelines will be central to advancing HGNN anomaly detection from research prototypes to robust, field-deployable cybersecurity systems.

\section{Conclusion}
This survey has provided a comprehensive review of heterogeneous graph neural network methods for anomaly detection, with a focus on node-level, edge-level, and subgraph-level tasks. We introduced a taxonomy that categorizes existing models based on their anomaly detection targets and learning paradigms, encompassing reconstruction-based, contrastive, attention-driven, and temporal models. We reviewed over 100+ studies and cited 89 representative papers to highlight how HGNNs effectively leverage heterogeneous graph structures, metapath-based semantics, and temporal dependencies to address complex anomaly detection scenarios.

A critical examination of benchmark datasets revealed both the strengths and limitations of existing resources. While widely used datasets such as Cora, Amazon, and CICIDS2017 have supported model evaluation, they often exhibit synthetic or sparse anomalies, limited heterogeneity, and inconsistent graph construction protocols. Future progress will rely on the creation of large-scale, realistic, and temporally annotated datasets that capture the dynamic nature of cybersecurity environments. Similarly, evaluation metrics such as AUROC, AUPRC, Precision@$K$, and NDCG should evolve toward standardized, task-aware protocols that incorporate operational factors such as detection latency, interpretability, and false-alert cost. Developing such unified benchmarks and evaluation suites will be essential for ensuring fair comparison, reproducibility, and meaningful advancement in HGNN-based anomaly detection.

Several persistent challenges emerged from this survey. First, the lack of standardized and richly annotated heterogeneous graph datasets remains a significant bottleneck, hindering fair comparisons and reproducibility. Second, most benchmarks are static, limiting the evaluation of dynamic or streaming HGNN models that are increasingly relevant in real-world applications such as cybersecurity and fraud detection. Third, evaluation practices often rely on a narrow set of metrics, which may not fully capture operational performance, especially in highly imbalanced settings.

Looking ahead, we identify three key directions for future research. First, there is a pressing need to develop benchmark suites that provide standardized graph schemas, temporal annotations, and clear anomaly ground truth across diverse domains. Second, more attention should be paid to robust evaluation frameworks, incorporating adaptive thresholding, uncertainty quantification, and cost-sensitive metrics to better reflect deployment realities. Third, advancing HGNN architectures that can jointly model structural, semantic, and temporal irregularities while also maintaining scalability and interpretability remains an open research frontier. By consolidating the current landscape of HGNN-based anomaly detection, this survey aims to serve as both a reference and a catalyst for future innovation in the field. We envision that advances in HGNNs, supported by standardized datasets and robust evaluation protocols, will pave the way toward operationally viable anomaly detection systems that are not only accurate but also scalable, interpretable, and resilient against evolving threats.

\end{document}